\documentclass[3p,times,twocolumn]{elsarticle}

\usepackage{ecrc}
\usepackage{epstopdf}

\volume{00}

\firstpage{1}

\journalname{Nuclear Physics B Proceedings Supplement}

\runauth{E.G. Berezhko}

\jid{nuphbp}

\usepackage{amssymb}

\usepackage[figuresright]{rotating}

\newcommand{\lsim}{\,\raisebox{0.2em}{$<$}\!\!\!\!\!
\raisebox{-0.25em}{$\sim$}\,}
\newcommand{\gr}{$\gamma$-ray \,}
\newcommand{\grs}{$\gamma$-rays \,}
\newcommand{\rxj}{RX~J1713.7-3946 }

\begin{document}

\begin{frontmatter}



\dochead{}

\title{Origin of Galactic Cosmic Rays from Supernova Remnants}


\author{E.G. Berezhko}

\ead{berezhko@ikfia.ysn.ru}

\address{Yu. G. Shafer Institute of Cosmophysical Research and 
Aeronomy SB RAS, 31 Lenin Ave., 677891 Yakutsk, Russia}

\begin{abstract}
  We analyze the results of recent measurements of Galactic cosmic ray (GCRs)
  energy spectra and the spectra of nonthermal emission from supernova remnants
  (SNRs) in order to determine their consistency with GCR origin in SNRs.  It
  is shown that  the measured primary and secondary CR nuclei energy
  spectra as well as the observed positron-to-electron ratio are consistent
  with the origin of GCRs up to the energy $10^{17}$~eV in SNRs.  Existing SNR
  emission data provide evidences for  efficient CR
  production in SNRs accompanied by significant magnetic field
  amplification. In some cases the nature of the detected \gr emission is
  difficult to determine because key SNR parameters are not known or poorly
  constrained.
\end{abstract}


\end{frontmatter}





\section{Introduction}

Supernova remnants (SNRs) are considered as a main cosmic ray (CR) source. They
are able to support  a constant density of  the Galactic cosmic ray (GCR)
population against loss by escape, nuclear interactions and ionization energy
loss.  The mechanical energy input to the Galaxy from each supernova (SN) is
about $10^{51}$ erg so that with a rate of about one every 30 years the total
mechanical power input from supernovae is of the order $10^{42}$ erg/s
(e.g. \citep{gs64,berezinetal90}).  Thus supernovae have enough power to drive
the GCR acceleration if there exists a mechanism for channeling about 10\% of
the mechanical energy into relativistic particles.

An appropriate acceleration mechanism is known since 1977 \citep{krym77}.
This is so called regular or diffusive shock acceleration process.  The strong
shock produced by high velocity ejecta expanding into the ambient medium pick
up a few particles from the plasma flowing into the shock fronts and accelerate
them to high energies.

The theory of particle acceleration by the strong shocks associated with SNRs
at present is sufficiently well developed and specific to allow quantitative
model calculations (e.g. see \citep{drury83, blei87, bk88, dm, ber08} for
reviews).  Theoretically progress has been due to the development of the
kinetic nonlinear theory of diffusive shock acceleration \citep{bek96, bv97,
  bv00a, zp09, kang10}.  The theory consistently includes the most relevant
physical factors, essential for SNR evolution and CR acceleration, and it is
able to make quantitative predictions of the expected properties of CRs
produced in SNRs and their nonthermal radiation.

There are also strong theoretical and observational reasons, that argue for a
significant amplification of the magnetic field as a result of the pressure
gradient of the accelerating CRs, exciting instabilities in the precursor of
the SNR shock. The most important consequence of magnetic field amplification
in SNRs is the substantial increase of the maximal energy of CRs, accelerated
by SN shocks, that presumably provides the formation of GCR spectrum inside
SNRs up to the energy $10^{17}$~eV.

Considerable progress have been achieved during the last years in experimental
determination of GCR spectra and spectra of nonthermal emission of SNRs. To
empirically confirm that the main part of GCRs indeed originates in SNRs one
has to check the consistency of the theoretical expectations with the observed
properties of GCRs and nonthermal emission of SNRs.

Here we analyze the existing data especially those obtained in recent
experiments PAMELA, Fermi and AMS-02 together with observational results of
nonthermal emission of SNRs in order to check their consistency with GCR origin
in SNRs.

\section{Production of  CRs in SNRs}
Acceleration of CRs in SNRs starts at some relatively low 
energy when some kind of suprathermal particles begin to cross the
SNR shock front. Any mechanism which supply suprathermal particles
into the shock acceleration is called injection.

Some small fraction of the
postshock thermal gas particle population are able to recross the shock
that means the beginning of their shock acceleration.
This is the most general and the most intense injection mechanism.
It occurs for all kind of ions and electrons existing
in the interstellar medium (ISM) and therefore it is
relevant for primary CRs only.
The corresponding injection rate is determined by the number of
particles involved into the acceleration from each medium volume
crossed the shock
and by the momentum of these particles \citep{bek96}:
\begin{equation}
N_\mathrm{inj}= \eta N_\mathrm{e1}, ~
p_\mathrm{inj}=\lambda m c_\mathrm{s2}.  
\label{eq1}
\end{equation} 
Here $N_\mathrm{e}$ is the number density of considered elements in ISM,
$c_\mathrm{s}$ is the sound speed, the subscripts 1(2) refer to the point just
ahead (behind) the shock.  Typical values of the dimensionless injection
parameters which provide CR production with required efficiency are $\eta =
3\times 10^{-4}$ and $\lambda =4$.  Since positrons, antiprotons and nuclei Li,
Be, B are not represented in the ISM secondary CRs can not be produced due to
such an injection.

Kinetic energy of all kind of GCR particles 
is considerably larger then the energy of gas particles injected
from the postshock thermal pool. Therefore
all GCRs which meet the expanding SNR shock are naturally involved
into the diffusive shock acceleration.
CR acceleration due to this  second relevant injection mechanism  is usually
called "reacceleration".

Since GCR energy spectra are relatively steep
and have a peak at kinetic energy
$\epsilon_{k} =\epsilon_\mathrm{GCR}\sim 1$~GeV  
their injection parameters can be represented in the form
\begin{equation}
N_\mathrm{inj}=N_\mathrm{GCR},~
 p_\mathrm{inj}= p_\mathrm{GCR},
\label{eq2}
\end{equation}
where $N_\mathrm{GCR}$ is the total number of GCR species per unit volume and 
$p_\mathrm{GCR}$
is their mean momentum, that corresponds to $\epsilon_\mathrm{GCR}$.
 
Primary nuclei during their acceleration inside SNRs produce secondary nuclei
in nuclear collisions with the background gas like GCRs do it in the Galactic
disk. Essential fraction of these already energetic particles has possibility
to be involved in further shock acceleration.  This is the third mechanism of
CR production inside SNRs.  For the first time it was studied
to describe the formation of secondary CR nuclei spectra \citep{bkpvz03}.

Secondaries with momentum $p$ are created throughout the remnant, everywhere
downstream and upstream of SNR shock up to the distances $d\sim
l_\mathrm{p}(p')$ of the order of the diffusive length $l_\mathrm{p}(p')$ of
their parent primary CRs with momentum $p'>p$.  Essential part of these
particles are naturally involving in the acceleration at SNR shock.  It
includes all the particles created upstream and the particles created
downstream at distances less than their diffusive length $l_\mathrm{s}(p)$ from
the shock front.  Since diffusive length $l\propto \kappa(p)\propto p$ is
increasing function of momentum for the Bohm type diffusion coefficient
$\kappa(p)$ which is realized during efficient CR acceleration in SNRs the
spectrum of secondary CRs for the first time intersecting the shock front is
very hard $N_\mathrm{inj}(p) \propto p^{-1}$ within wide range of their momenta
$p$.  This makes the secondary particle spectra $N_\mathrm{s}(p,t)$, produced in
SNR, harder compared with the spectra of primaries $N_\mathrm{p}(p,t)$.

The SNR efficiently accelerates CRs up to some maximal age $T_\mathrm{SN}{\tiny
  \approx} {\bf \sim} 10^5$~yr when SNR release all previously accelerated CRs,
primaries and secondaries, with the energy spectra
$N_\mathrm{p}(\epsilon_{k},T_\mathrm{SN})$ and
$N_\mathrm{s}(\epsilon_{k},T_\mathrm{SN})$ respectively, into surrounding ISM.
Here $\epsilon_\mathrm{k}$ is the particle kinetic energy.  These CRs released
from SNRs together with secondary CRs produced in ISM form the total secondary
$n_\mathrm{s}(\epsilon_\mathrm{k})$ and primary
$n_\mathrm{p}(\epsilon_\mathrm{k})$ CR populations. At sufficiently high
energies the s/p ratio of nuclear component within simple leaky box model is
given by the expression \citep{bkpvz03}
\begin{equation}
\frac{n_\mathrm{s}}{n_\mathrm{p}}=\frac{n'_\mathrm{s}}{n_\mathrm{p}}+
\frac{N_\mathrm{s}}{N_\mathrm{p}},
\label{eq5}
\end{equation}
where 
$n'_\mathrm{s}(\epsilon_\mathrm{k})$ represents the spectrum of 
secondaries produced in nuclear collisions of
primary CRs within the Galactic disk.
It
is approximately given by the expression \citep{bkpvz03}
${n'_\mathrm{s}}/{n_\mathrm{p}} = \sigma x/m_\mathrm{p}$, where 
$x=\rho v\tau_\mathrm{esc}$ 
is the escape length which is the mean matter thickness traversed by GCRs in 
the course of their random walk in the Galaxy,
$\rho$ is the ISM gas density, 
$\tau_\mathrm{esc}(\epsilon_\mathrm{k})$ 
is the CR escape time from the Galaxy, $m_\mathrm{p}$ is the proton mass,
$\sigma$ is the crossection of secondary CRs production.

At sufficiently high energies the s/p ratio $n_\mathrm{s}/n_\mathrm{p} \approx
N_\mathrm{s}/N_\mathrm{p}$ is determined by the ratio
$N_\mathrm{s}/N_\mathrm{p}$ produced in the SNRs independently on the
propagation model. The same is  true for positron to
electron ratio even though electrons are not the parent particles for
positrons.

The rigidity spectrum of primary CRs accelerated by strong SNR shock
within the test particle limit has a pure power law form
$N_p \propto R^{-\gamma}$, which extends up to some maximal
value $R_\mathrm{max}$, with power law index $\gamma\approx 2$.
Here  $R= pc/(Ze)$  and $Z$ are the 
rigidity and the charge number of CR particle respectively,
$c$ is the speed of light, $e$ is the proton charge.
In the case of efficient CR injection/acceleration which
takes place at the expected proton injection rates $\eta \gg 10^{-5}$
the shock transition becomes modified due to CR backreaction.
The spectrum of CRs accelerated by the modified shock is not a pure power law,
it has a concave shape so that it becomes progressively flatter at higher rigidity: 
at low energies $R\ll R_\mathrm{i}$ $N_p \propto R^{-\gamma_\mathrm{l}}$
with $\gamma_\mathrm{1} >2$ whereas at high energies 
$R_\mathrm{i} \ll R< R_\mathrm{max}$ $N_\mathrm{p} \propto R^{-\gamma_\mathrm{h}}$
with $\gamma_\mathrm{h} <2$. The deviation from the pure power law spectrum
depends on the shock strength so that at early SNR evolutionary epoch
$\gamma_\mathrm{1}$ can be as large as $\gamma_\mathrm{1} = 3$ and $\gamma_\mathrm{h}$ can be as
small as $\gamma_h =1.8$. The transition region from the steepest part
of CR spectrum to its hardest part takes place at 
$R_\mathrm{i}\sim 100m_\mathrm{p}c^2/e$ or at the energy 
$\epsilon_\mathrm{i}\sim 100Zm_\mathrm{p}c^2$.

Since the CR residence time within the Galactic confinement volume (or the mean
escape time of CRs from this volume) is expected to have power law dependence
on CR rigidity $\tau_\mathrm{esc}\propto R^{-\mu}$ the spectrum of primary GCR
spectrum $n_\mathrm{p} \propto \tau_\mathrm{esc}N_\mathrm{p}$ remains concave.

The spectrum of secondary GCRs, $n_\mathrm{s}(R)$, consists of two parts. Its low
energy part produced in nuclear collisions of primary CRs with ISM nuclei is
very steep $n_\mathrm{s}(R) = n'_\mathrm{s}(R)\propto
\tau_\mathrm{esc}(R)n_\mathrm{p}(R)$ whereas its high energy part, dominated by
the SNR contribution, is expected to be even harder then the primary GCR
spectrum. Therefore the secondary GCR spectra are expected to be considerably
more concave compared with the primary GCR spectra.  As a consequence the
rigidity dependence of s/p ratio should have also the concave form.

There are strong theoretical reasons \citep{lb00,belll01,bell04,caprioli14} and
observations of correspondingly sharp X-ray synchrotron filaments in  e.g.
SN 1006 \citep{bamba03,long03}, that argue for a significant amplification of
the magnetic field as a result of the pressure gradient of the accelerating
CRs, exciting instabilities in the precursor of the SNR shock \citep{bkv03,
  bv04}. However, as yet there does not exist a convincing dynamical equation
for the resulting field amplitudes (for hybrid simulations and a most recent
argument regarding large-scale field generation,
e.g.\citep{reville12}). Therefore, field amplification is not directly
contained in the form of a kinetic wave equation in the existing nonlinear
theory.  The effective upstream field amplitude in the precursor $B_0(t)$ at
the current epoch $t$ is rather determined as a result of the best fit of the
calculated to the measured spatially integrated synchrotron spectrum.

The strength of amplified upstream magnetic field $B_0(t)$ varies
according to the  empirical relation \citep{vbk05}
\begin{equation}
B_0^2/(8\pi)=2.5\times 10^{-4}\rho V_\mathrm{s}^2.
\end{equation}

The most important consequence of magnetic field amplification in SNRs is the
substantial increase of the maximum energy of CRs
$\epsilon_\mathrm{max}$, accelerated by SN shocks. 
Cutoff energy of CRs accelerated at any given SNR evolutionary stage is
determined by geometrical factors and its value can be presented in the form
\citep{ber96}
\begin{equation}
\epsilon_\mathrm{m}\propto  ZB_0R_\mathrm{s}V_\mathrm{s},
\end{equation}
where $R_\mathrm{s}$ and $V_\mathrm{s}$ are the shock radius and speed respectively.
Therefore the upper (cutoff) CR energy during the Sedov evolutionary phase
where the main part of CRs are produced varies as
$\epsilon_\mathrm{m} \propto t^{-4/5}$.
The maximal CR energy 
$\epsilon_\mathrm{max} =\mbox{max} \{\epsilon_\mathrm{m}(t)\} \propto V_0 R_0 B_0(t_0)$
is produced at the beginning of the 
Sedov phase $t\approx t_0$ \citep{ber96}.
Magnetic field amplification in typical SNR
provides maximal energy for protons  
$\epsilon_\mathrm{max}\approx 3\times 10^6m_\mathrm{p}c^2$ that is required 
condition to reproduce
the spectrum of GCRs up to the knee energy \citep{bv07}. 

The most general properties of CR acceleration in SNRs 
-- concave CR rigidity spectra, concave s/p ratios, time dependence
of the CR cutoff energy $\epsilon_\mathrm{m}(t)$ -- briefly described
in this section play primarily important role when
one analyze the consistency of existing data with
GCR origin in SNRs.

\section{Primary GCRs}

\begin{figure}[t]
\includegraphics[width=0.47\textwidth]{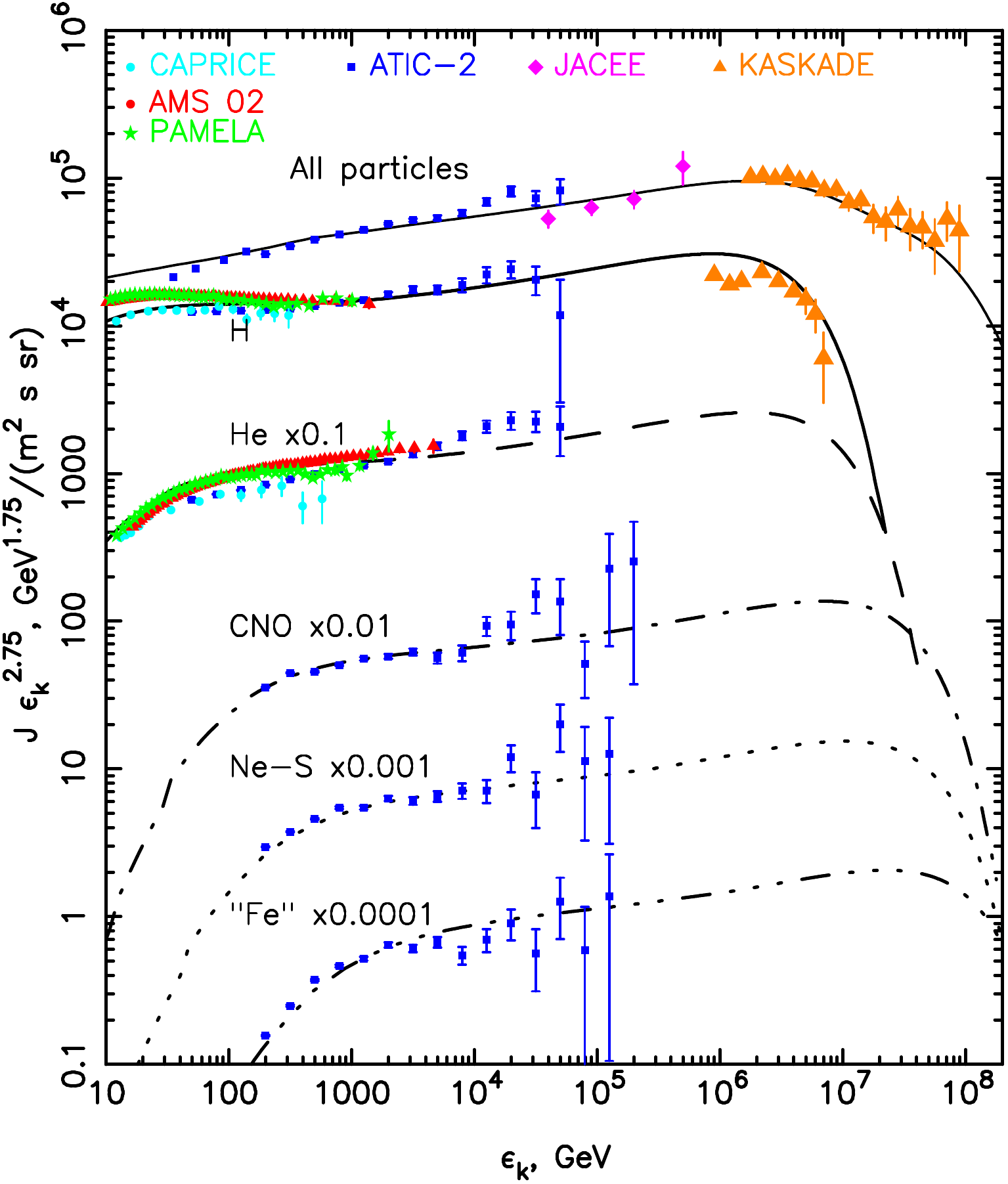}
\caption{GCR intensities at the Solar system 
as a function of kinetic energy.
Experimental data obtained in the CAPRICE \citep{caprice}, AMS 02 \citep{ams2}
PAMELA \citep{pamelaCR},
ATIC-2 \citep{atic2}, JACEE \citep{jacee} and KASCADE \citep{kascade}
experiments are
shown as well.The calculated curves were normalized to the ATIC-2
data at $\epsilon_\mathrm{k}>1$~TeV}
\label{Fig1}
\end{figure}

The intensities of protons, Helium, three groups of heavier nuclei, and "All
particles" as a function of kinetic energy calculated within kinetic nonlinear
theory \citep{bv07} are presented in Fig.~\ref{Fig1} (see also \citep{pzs10}
for a similar consideration).  The values $E_\mathrm{SN}=10^{51}$~erg for the
explosion energy, $M_\mathrm{ej}=1.4 M_{\odot}$ for the ejecta mass,
$T_\mathrm{SN}= 10^5$~yr for the active SNR age and $\tau_\mathrm{esc} \propto
R^{-\mu}$, with $\mu = 0.75$ for the escape time were used here. One can see
that the theory fits the existing data in a satisfactory way up to the energy
$\epsilon_\mathrm{k} \approx 10^{17}$~eV. This is especially 
  true if one considers the difference between the CAPRICE and ATIC-2
experiments from one side and AMS-02 and PAMELA experiments from the other side
as a measure of actual experimental uncertainty in these experiments.
Therefore it is not so clear whether the Helium spectrum is indeed so
noticeably harder than the proton spectrum as measured in the ATIC-2 balloon
experiment.

The difficulty for the present theory is the fact that, in
order to get consistency with the observed GCR spectra, a strongly energy
dependent escape time $\tau_\mathrm{esc}\propto R^{-\mu}$ with $\mu=0.75$ is needed,
which is beyond the experimentally determined interval
$\mu=0.3-0.7$ \citep{berezinetal90}. It is not clear at the moment whether or
not CR escape from the Galaxy can be so strongly energy dependent. 
%

The calculation predicts concave spectra
at energies $\epsilon_\mathrm{k}>10Z$~GeV
for individual CR species and there is some experimental evidence that the
actual proton GCR spectrum is indeed concave.  It is difficult to conclude
whether Helium and heavier nuclei energy spectra have also concave shape.  One
needs more precise measurements at $\epsilon>10^4$~GeV in order to be able to
arrive at a more stringent conclusion. At the same time the calculated
all-particle spectrum has almost a pure power law form up to the knee energy
(see Fig.~\ref{Fig1}).

According to Fig.~\ref{Fig1} the knee in the observed all-particle GCR spectrum
has to be attributed to the maximum energy of protons, produced in SNRs. The
steepening of the all particle GCR spectrum above the knee energy $3\times
10^{15}$~eV is a result of the progressively decreasing contribution of light
CR nuclei with increasing energy. Such a scenario is confirmed by the KASCADE
experiment which shows relatively sharp cutoffs of the spectra of various GCR
species at energies $\epsilon_\mathrm{max}\approx 3Z\times10^{15}$~eV
\citep{kascade}, so that at energy $\epsilon \sim 10^{17}$~eV the GCR spectrum
is expected to be dominated by the contribution from the iron nuclei.

One can conclude that the existing measurements of primary GCR nuclei
are generally consistent with their origin in SNRs.
%
\begin{figure}
\includegraphics[width=0.47\textwidth]{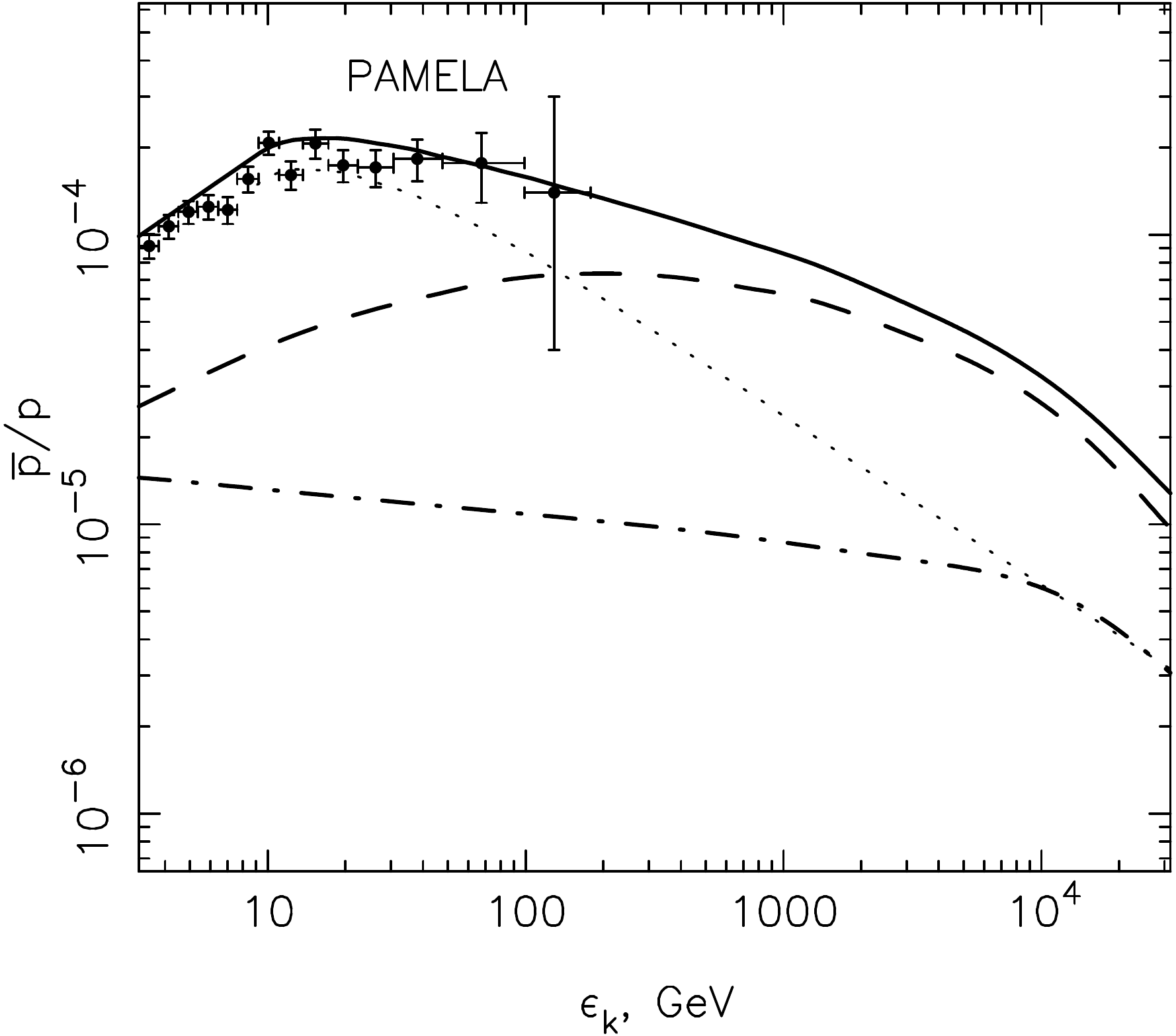}
\caption{\label{Fig2} Calculated antiproton-to-proton ratio as a function of
  energy \citep{bk14} together with PAMELA \citep{pamela10} data.  Dotted and
  dashed lines correspond  to spectra of antiprotons created in p-p
  collisions in ISM \citep{donato01} and inside SNR respectively, the
  dash-dotted line corresponds to the spectrum of antiprotons reaccelerated in
  SNRs, the solid line represents the sum of contributions of all these
  processes.  }
\end{figure}
%
%

\section{Secondary GCRs}

The antiproton-to-proton ratio $\bar p/p=n_{\bar p}(\epsilon_\mathrm{k})/
n_{p}(\epsilon_\mathrm{k})$ as a function of kinetic energy, calculated within
kinetic nonlinear model \citep{bk14}, together with PAMELA data are shown in
Fig.~\ref{Fig2}.  It is seen that antiprotons  for energies
$\epsilon_\mathrm{k}<10$~GeV are produced in SNRs equally effectively by 
  two mechanisms inside SNRs  whereas  for $\epsilon_\mathrm{k}>10$~GeV the creation of antiprotons in p-p
collisions and their subsequent acceleration  -- both inside the SNRs --
 become dominant.  In total the antiproton production in
SNRs makes the energy dependence of $\bar p/p$ considerably more flatter so
that at $\epsilon_\mathrm{k}\sim 10^3$~GeV the ratio becomes larger by a factor
of about five.  Within the energy range 30~GeV~$<\epsilon_\mathrm{k}< 3000$~GeV
the energy dependence of the ratio $\bar p/p\propto
\epsilon_\mathrm{k}^{-0.25}$ is expected to be very flat.

PAMELA data, which well agree with the calculation \citep{bk14} (see also
\citep{BlasiSer09}) , within the energy range 10~GeV~$<\epsilon_\mathrm{k}<
100$~GeV provide the evidence that the actual ratio $\bar p/p$ is indeed  
flatter than it is expected if antiprotons are created in the ISM only.

\begin{figure}
\includegraphics[width=0.47\textwidth]{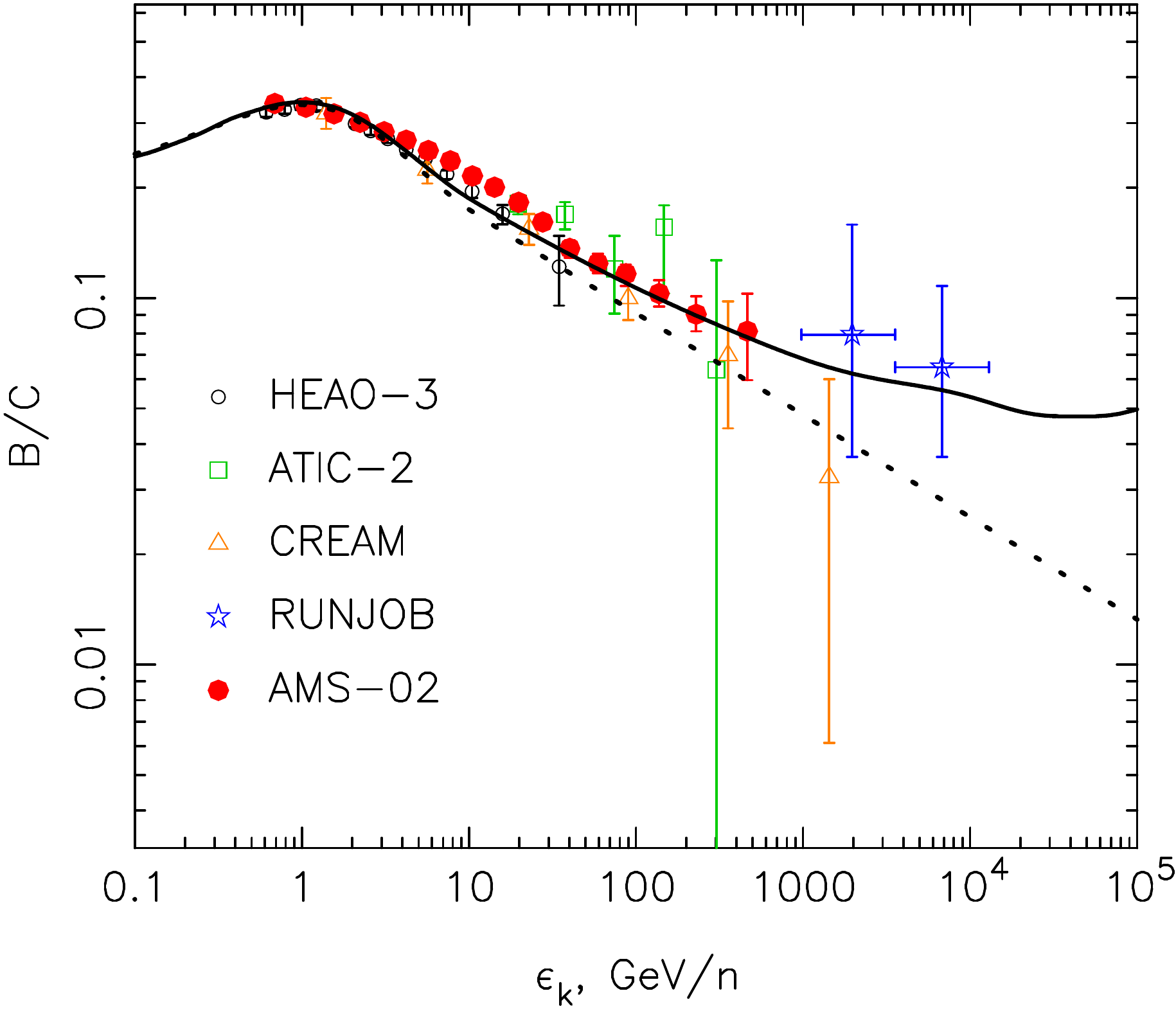}
\caption{\label{Fig3}
Calculated boron-to-carbon  ratio as a function of 
kinetic energy per nucleus
together with the results of
HEAO-3 \citep{heao3}, ATIC-2 \citep{atic2bc}, CREAM \citep{cream},
RUNJOB\citep{runjob} and AMS-02 \citep{ams02} experiments.
Dotted line corresponds to spectrum of boron created  in nuclear collisions
in ISM \citep{bkpvz03}, solid line represents the boron spectrum which
includes the contribution of SNRs. 
 }
\end{figure}

\begin{figure}
\includegraphics[width=0.47\textwidth]{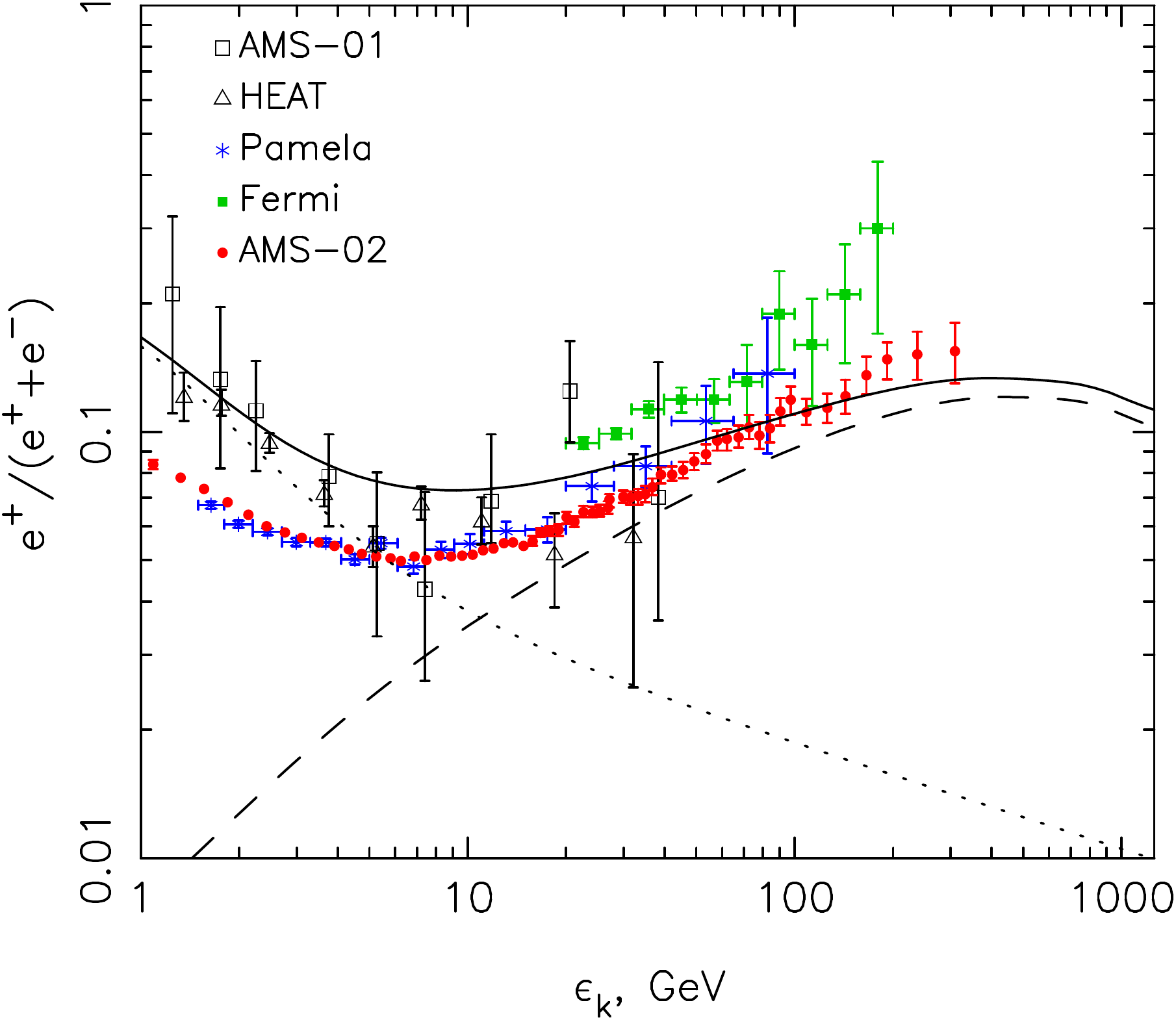}
\caption{\label{Fig4} Calculated positron to electron plus positron ratio as a
  function of kinetic energy together with the results of AMS-01 \citep{ams01},
  HEAT \citep{heat}, PAMELA \citep{pamela09}, Fermi \citep{fermi12} and AMS-02
  \citep{ams13} experiments.  Dotted \citep{ms98} and dashed lines correspond
  to positrons produced in ISM and inside SNRs. respectively;  the solid
  line represents the sum of contributions.}
\end{figure}

In order to check the consistency of other types of secondary CR production the
boron-to-carbon (B/C) and positron to electron plus positron ($e^+/(e^+ +e^-)$)
ratios calculated within the same model and compared in Fig.~\ref{Fig3} and
\ref{Fig4} respectively with the existing experimental data \citep{bk14}.

Due to the boron production in SNRs the expected B/C ratio undergoes
considerable flattening which starts at energy $\epsilon_\mathrm{k}\approx
100$~GeV/n.  As one can see in Fig.~\ref{Fig3} this is consistent with the
measurements recently performed in RUNJOB balloon \citep{runjob} and AMS-02
space \citep{ams02} experiments even though for more strict conclusion one
needs the measurements with higher statistics at energies above 1~TeV/n.

%

As it is seen in Fig.~\ref{Fig4} that the expected ratio $e^+/(e^+ +e^-)$ due
to SNRs contribution very well corresponds to the growing part of this ratio
detected in PAMELA \citep{pamela09}, Fermi \citep{fermi12} and AMS-02
\citep{ams13} experiments at energy $\epsilon_\mathrm{k} > 10$~GeV.
Quantitatively there is a good agreement between calculation and AMS-02 data at
energy $\epsilon_\mathrm{k} > 30$~GeV.  Note, that the predicted overproduction
of positrons at $\epsilon_\mathrm{k} < 20$~GeV is  due to the
adopted model of CR propagation inside the Galaxy \citep{ms98}.

The calculated SNR contribution to the secondary CR spectra represents the
component which is unavoidably expected if SNRs are the main source of GCRs.
Comparison with the existing data leads to the conclusion that the observed high
energy excess of secondary nuclei and positrons can be produced in Galactic
SNRs. This enables to expect a similar excess in the antiproton energy
spectrum. The data expected very soon from AMS-02 experiment will make it clear
whether the actual ratio $\bar p/p$ is indeed not less flat at energies
$\epsilon_{k}> 10$~GeV than we predict.

\section{Nonthermal emission of Galactic SNRs}

Direct information about the high-energy CR population in SNRs can be obtained
from the observations of the nonthermal emission produced by accelerated CRs in
SNRs. The electron CR component is evident in a wide wave length range by the
radiation that they produce in SNRs, from radio to $\gamma$-ray emission,
whereas in the case of the nuclear CR component a $\gamma$-ray detection is the
only possibility to find it.  If this nuclear component is strongly enhanced
inside SNRs then through inelastic nuclear collisions, leading to pion
production and subsequent decay, $\gamma$-rays will be produced at the
detectable level (e.g.\citep{druryetal94,bv97,bv00a}).

There are a number of physical parameters, whose values are needed to be known
to describe the observed properties of young SNRs. SN explosion energy
$E_\mathrm{sn}$, ejecta mass $M_\mathrm{ej}$, ISM number density $N_\mathrm{H}$
directly determine the SNR dynamics, the observed size $R_\mathrm{s}$ and speed
of SN shock $V_\mathrm{s}$. Therefore their values are mainly determined from
the fit of $R_\mathrm{s}(t)$ and $V_\mathrm{s}(t)$ if the SNR age $t$ and the
SNR distance $d$ are known.  At the same time there are two other physical
parameters which essentially influence the efficiency of the diffusive shock
acceleration and its final significance.

The first one is the proton injection rate $\eta$.  There is a good
semi-quantitative understanding of the injection rate of nuclear particles into
the diffusive shock acceleration process. 
The actual value of $\eta$ can be extracted from the fit of the observed
SNR synchrotron spectrum, whose shape is sensitive to the shock modification
due to the CR backreaction that in turn strongly depends on $\eta$.

The magnetic field strength is the second important parameter.  It determines
the  maximum energy of nuclear CRs which could be achieved
during the SNR evolution $\epsilon_\mathrm{max}\propto B_0$.  Radio
observations of synchrotron emission are powerful probes of magnetic fields and
electron distributions in SNRs. Electrons with a power-law energy spectrum
$N_\mathrm{e}(\epsilon)= A_\mathrm{e} \epsilon ^{-\gamma}$ produce the
synchrotron flux $S_{\nu}\propto A_\mathrm{e} B_\mathrm{d}^{(\gamma+1)/2}\nu
^{-\alpha}$ with the spectral index $\alpha=(\gamma-1)/2$.  Here
$B_\mathrm{d}\approx \sigma B_0$ is the downstream magnetic field strength,
$\sigma$ is the shock compression ratio.  In the test-particle limit power-law
index $\gamma=2$ and therefore $\alpha=0.5$.  Values $\alpha>0.5$ observed in
young SNRs require concave electron spectrum (progressive hardening to higher
energies) with $\gamma> 2$ at $\epsilon < 1$~GeV, as predicted by nonlinear
shock acceleration models. The required shock modification is
produced by the proton CR component. Therefore the fit of the spectral shape of
the radio emission makes it possible to determine the actual proton injection
rate $\eta$.  In addition, since the synchrotron emission at frequency $\nu$
are mainly produced by electrons of energy $ \propto \sqrt{\nu/B_\mathrm{d}}$ one needs
not only efficient CR acceleration with subsequent shock modification but also
a high magnetic field $B_\mathrm{d}\gg 10$~$\mu$G in the acceleration region in
order to have $\alpha>0.5$ at radio frequencies.

All energy of
the electrons at high energies $\epsilon>\epsilon_\mathrm{l}\propto
B_\mathrm{d}^{-2}$, where synchrotron losses are significant, is rapidly and
completely transformed into the synchrotron emission
independently on the magnetic field value.
The synchrotron flux in this X-ray  range
is determined by the number of injected/accelerated electrons, 
or by the amplitude $A_\mathrm{e} \propto \eta K_\mathrm{ep}$. Therefore
the fit of X-ray synchrotron spectrum at known proton injection rate gives 
the estimate of electron to proton ratio $K_\mathrm{ep}$.
At known $A_\mathrm{e}$ synchrotron radio emission
$S_{\nu}\propto  B_\mathrm{d}^{(\gamma+1)/2}\nu ^{-\alpha}$ depends
on the interior magnetic field value $B_\mathrm{d}$. Therefore
the fit of radio emission provides the estimate of
downstream magnetic field strength $B_\mathrm{d}$.

In addition, since for given synchrotron flux $S_{\nu}$ required number of electrons
is determined by the amplitude
$A_\mathrm{e}\propto B_\mathrm{d}^{-(\gamma+1)/2}$, 
the relative role of electrons in high-energy $\gamma$-ray production is
expected to be low in young SNRs where magnetic field is strongly amplified.

Recent observations with the Chandra and XMM-Newton X-ray space telescopes
have confirmed earlier detections of nonthermal continuum emission in hard
X-rays from young shell-type SNRs. With Chandra it
became even possible to resolve spatial scales down to the arcsec
extension of individual dynamical structures like shocks
\citep{bamba03,long03,vl03}.
The filamentary hard X-ray structure is the result of strong synchrotron losses
of the emitting multi-TeV electrons in amplified magnetic fields downstream of
the outer accelerating SNR shock \citep{bkv03,bv04,vbk05}.  This provides an
unique possibility for the determination of the SNR magnetic field strength,
based on the measured filaments width \citep{bv04}.  These effective magnetic
fields turned out to be exactly the same as extracted from the fit of the
overall synchrotron spectrum.

In oder to perform the detail comparison of theoretical expectation with the
experiment one needs sufficient number 
of individual SNRs with known values of relevant
physical parameters such as the age, distance, ISM density.
Unfortunately the number of such SNRs are very limited, especially those which
are seen in all wavelength from radio to \gr. It is why
every new experimentally established property of nonthermal emission of
SNR represents considerable interest for theoretical
analysis. Here we analyze the properties of the nonthermal emission
from a number of young Galactic SNRs, which are detected \gr sources,
in order to find evidence for efficient CR acceleration
consistent with GCR origin in SNRs.

\subsection{Supernova remnant SN~1006}
Historical remnant SN~1006 is the most appropriate SNR for studying the
properties of nonthermal emission: the distance $d=2.2$~kpc was determined
using optical measurements with relatively high precision and all other
astronomical parameters are quite well known (see \citep{bkv12} and reference
there).

As a type Ia supernova SN~1006 presumably expands into a uniform ISM, ejecting
roughly a Chandrasekhar mass $M_\mathrm{ej}=1.4 M_{\odot}$.  The values of
several scalar key parameters (proton injection rate $\eta$, electron to proton
ratio $K_\mathrm{ep}$ and the interior magnetic field strength $B_\mathrm{d}$)
can be determined from a fit of the solutions that contain these parameters to
the observed spatially integrated synchrotron emission data.  The parameter
values for SN~1006, evaluated in this way \citep{bkv09}, agree very well with
the Chandra measurements of the X-ray synchrotron filaments.

In order to explain the detailed \gr spectrum, the values of the 
hydrodynamic supernova explosion energy $E_\mathrm{sn}= 2.4\times
10^{51}$~erg and $E_\mathrm{sn}= 1.9\times 10^{51}$~erg are taken to fit the
observed shock size $R_\mathrm{s}=9.5\pm0.35$~pc and shock speed
$V_\mathrm{s}=4500\pm 1300$~km/s  at the current
epoch $t_{SN}\approx 10^3$~yr for the ISM hydrogen number densities
$N_\mathrm{H}=0.08~\mbox{cm}^{-3}$ and $N_\mathrm{H}=0.05~\mbox{cm}^{-3}$,
respectively. These densities are consistent with the observed level of the \gr
emission.  Note that the best fit value of the upstream magnetic
field strength $B_0=30$~$\mu$G is almost non-sensitive to $N_\mathrm{H}$. 

\begin{figure}
\includegraphics[width=0.47\textwidth]{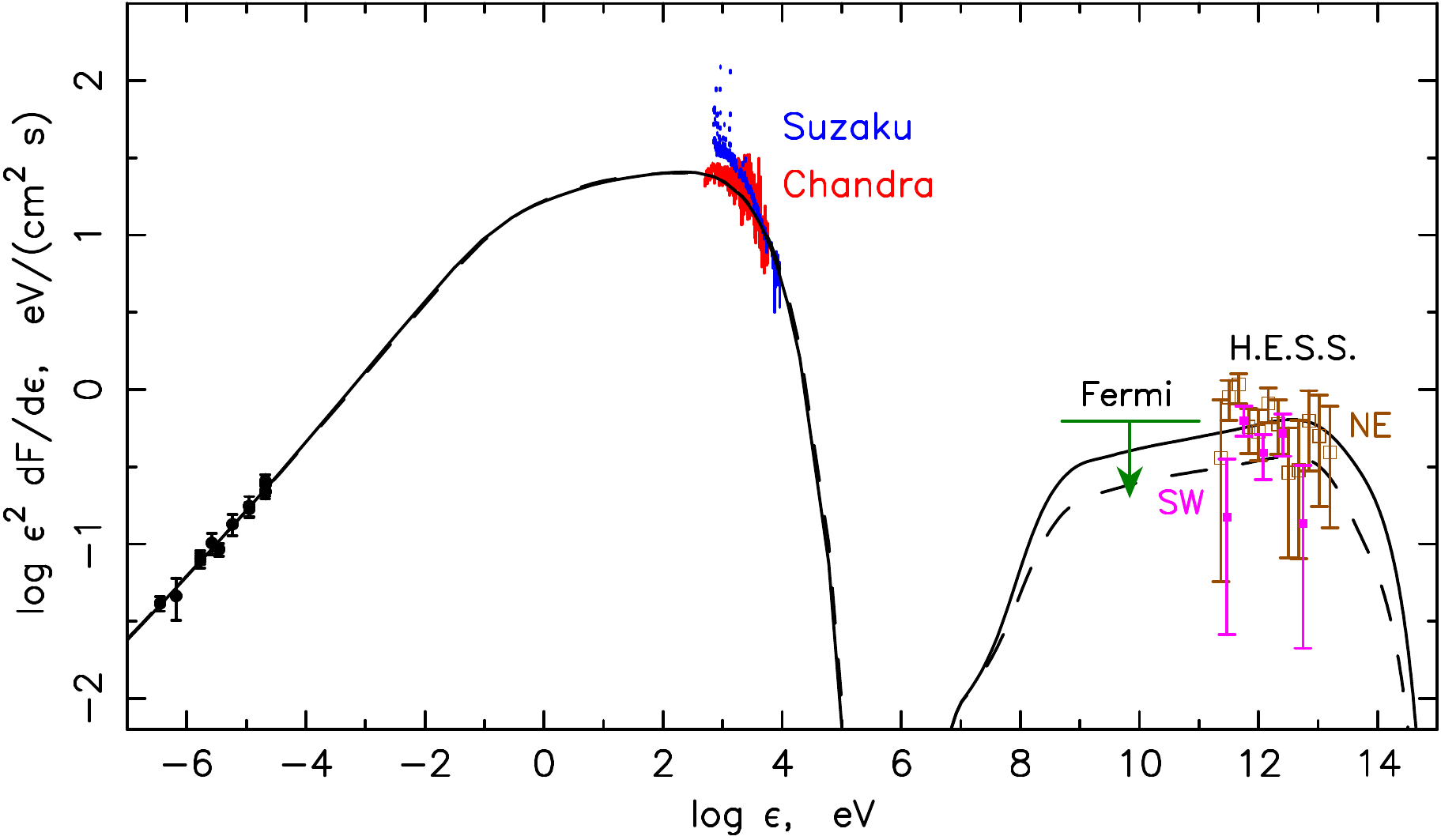}
\caption{Spatially integrated spectral energy distribution of SN~1006.  The
  observed radio spectrum \citet{allen04}, Chandra \citep{allen04} and Suzaku
  \citep{bamba08} X-ray spectra together with \gr H.E.S.S. data
  \cite{sn1006hess10} for the NE region and the SW region are shown as well.
  Both these H.E.S.S. data sets have been multiplied by a factor of 2, in order
  to make them comparable to the global theoretical spectra (solid and
    dashed curves). The Fermi upper limit of the GeV-emission is shown as well
  \citep{sn1006GeV}.  }
\label{Fig5}
\end{figure}

Fig.~\ref{Fig5} illustrates the consistency of the calculated synchrotron and
\gr spectra with the observed spatially integrated spectra \citep{bkv12}.  The
H.E.S.S. data \citep{sn1006hess10} for the NE and SW limbs, respectively, have
been multiplied by a factor of 2, in order facilitate comparison with the full
deduced \gr flux.  As can be seen from Fig.~\ref{Fig5}, the calculated
synchrotron spectrum fits the observations both in the radio and the X-ray
ranges very well.  Note that the radio spectral index value $\alpha =0.57$ is a
clear indication of the modified SNR shock.

The only important parameter which can not be determined from the analysis of
the synchrotron emission data is the external gas number density
$N_\mathrm{H}$: Fig.~\ref{Fig5} shows that the spectrum of synchrotron emission
is almost non-sensitive to the ambient gas density.  Consequently numerical
solutions have been calculated for the pair of values
$N_\mathrm{H}=0.08~\mbox{cm}^{-3}$ (solid curve) and
$N_\mathrm{H}=0.05~\mbox{cm}^{-3}$ (dashed curve) which appear to bracket
the density range consistent with the H.E.S.S. \gr measurements.
%

As a result the flux of TeV emission detected by H.E.S.S. is predicted to be
consistent with an ISM hydrogen number density $ N_\mathrm{H}\approx
0.06~\mbox{cm}^{-3}$, consistent with the expectation from X-ray
measurements. 

Above 10 GeV the theoretical spectral energy
flux can be  approximated analytically as a slowly
rising power law with an exponential cutoff
\[ 
dF_{\gamma}/d\epsilon_{\gamma}\propto \epsilon_{\gamma}^{-1.9}
\exp{(-\epsilon_{\gamma}/\epsilon_\mathrm{\gamma m})},
\] with cutoff energies $\epsilon_\mathrm{\gamma m}=30$~TeV and 37~TeV for
$N_\mathrm{H}=0.05~\mbox{cm}^{-3}$ and $N_\mathrm{H}=0.08~\mbox{cm}^{-3}$,
respectively.  These two analytical curves, representing the best fit to the
H.E.S.S.  data, within 4\%, respectively coincide with the calculated curves
presented in Fig.~\ref{Fig5}.

The Fermi upper limit at GeV energies \citep{sn1006GeV} is consistent with such
a hard \gr spectrum.

The cutoff energy value $\epsilon_\mathrm{\gamma m}=30$~TeV is unexpectedly low
since the overall proton power-law spectrum extends up to a proton energy
$\epsilon = \epsilon_m \approx 10^{15}$~eV \citep{bkv09}. The reason is the
progressive decrease of overlap between the radial gas density profile and the
radial profile of the proton distribution function at energies $\epsilon >
10^{14}$~eV.  This leads to a lower cutoff of the \gr spectrum
$\epsilon_\mathrm{\gamma m} \ll 0.1 \epsilon_\mathrm{m}$ compared with what one
would expect in the simple case of energy-independent overlap of CR protons
with the gas distribution.

This spread of CRs into the upstream region $r>R_\mathrm{s}$ which, for the
highest energies $\epsilon \sim \epsilon_\mathrm{m}(t)$, is faster than the
shock expansion $R_\mathrm{s}(t)$ and represents the diffusive escape of CRs
from the expanding SNR, as it was predicted \citep{bk88}. Such a CR escape is
expected in particular in the Sedov phase when the maximal energy
$\epsilon_\mathrm{m}(t)$ of CRs, accelerated in the given evolutionary phase,
decreases with time, because the value of $\epsilon_\mathrm{m}$ is shock-size
limited (see \citep{ber96} for details) rather than time limited. At any given
phase $t>t_0$ (where $t_0$ is the sweep-up time) efficient CR acceleration at
the SN shock takes place only for energies $\epsilon\le\epsilon_\mathrm{m}(t)$,
whereas for CRs with energies $\epsilon_\mathrm{m}(t)<\epsilon
<\epsilon_\mathrm{m}(t_0)$, produced during earlier times, the acceleration
process becomes inefficient and these CRs expand into the upstream region. CR
escape is relatively slow if Bohm diffusion in the amplified magnetic field
$B_0$ takes place everywhere upstream for any CR energy. In reality this is
expected to be true only for CRs with energies $\epsilon
<\epsilon_\mathrm{m}(t)$ in the vicinity of the shock, where these CRs produce
significant magnetic field amplification. At large distances upstream of the
shock, $r-R_\mathrm{s}\gg 0.1R_\mathrm{s}$, or/and for higher CR energies
$\epsilon >\epsilon_\mathrm{m}(t)$, CR diffusion  should be
much faster than Bohm diffusion. Therefore, in actual SNRs CR escape is
expected to be faster and more intense than the present model predicts (see
\citep{pz03} for more detailed considerations).  Since SN~1006 is only at the
very beginning of the Sedov phase this underestimate of the magnitude of escape
is not very critical however.

The sum of all above results suggests the conclusion that SN~1006  is a source with a
high efficiency of nuclear CR production, required for the GCR sources,
both in flux as well as in cutoff energy.

\subsection{Tycho's supernova remnant}

The kinetic nonlinear model has been used
in detail to investigate Tycho's SNR as the remnant of a type Ia SN in a
homogeneous ISM, in order to compare the results with existing data
\citep{vbk08}. It was argued that consistency of the standard value of stellar ejecta
mass $M_\mathrm{ej}=1.4M_{\odot}$ and a total hydrodynamical explosion energy
$E_\mathrm{sn}=1.2\times 10^{51}$~erg \citep{bad06} with the gas dynamics,
acceleration theory and the existing \gr measurements required the source
distance $d$ to exceed $3.3$~kpc in order to be consistent with the existing
HEGRA upper limit for TeV \gr emission. The corresponding ambient gas number
density $N_\mathrm{g}=1.4N_\mathrm{H}$  
had then to be lower than $0.4$~cm$^{-3}$. On the other hand, the rather
low distance estimates from independent measurements together with internal
consistency arguments of the theoretical model made it even more
likely that the actual \gr flux from Tycho is ``only slightly' below the
HEGRA upper limit. The strong magnetic field amplification produced by
accelerated CRs implied a mean field strength of $\approx 400~\mu$G and
as such implied in addition that the \gr flux is hadronically dominated.
The shock was modified with an overall
compression ratio $\sigma \approx 5.2$ and a subshock compression ratio
$\sigma_\mathrm{s} \approx 3.6$; the latter is consistent with the observed
radio index $\alpha\approx 0.61$ \citep{rey92}. 

\begin{figure}
\includegraphics[width=0.47\textwidth]{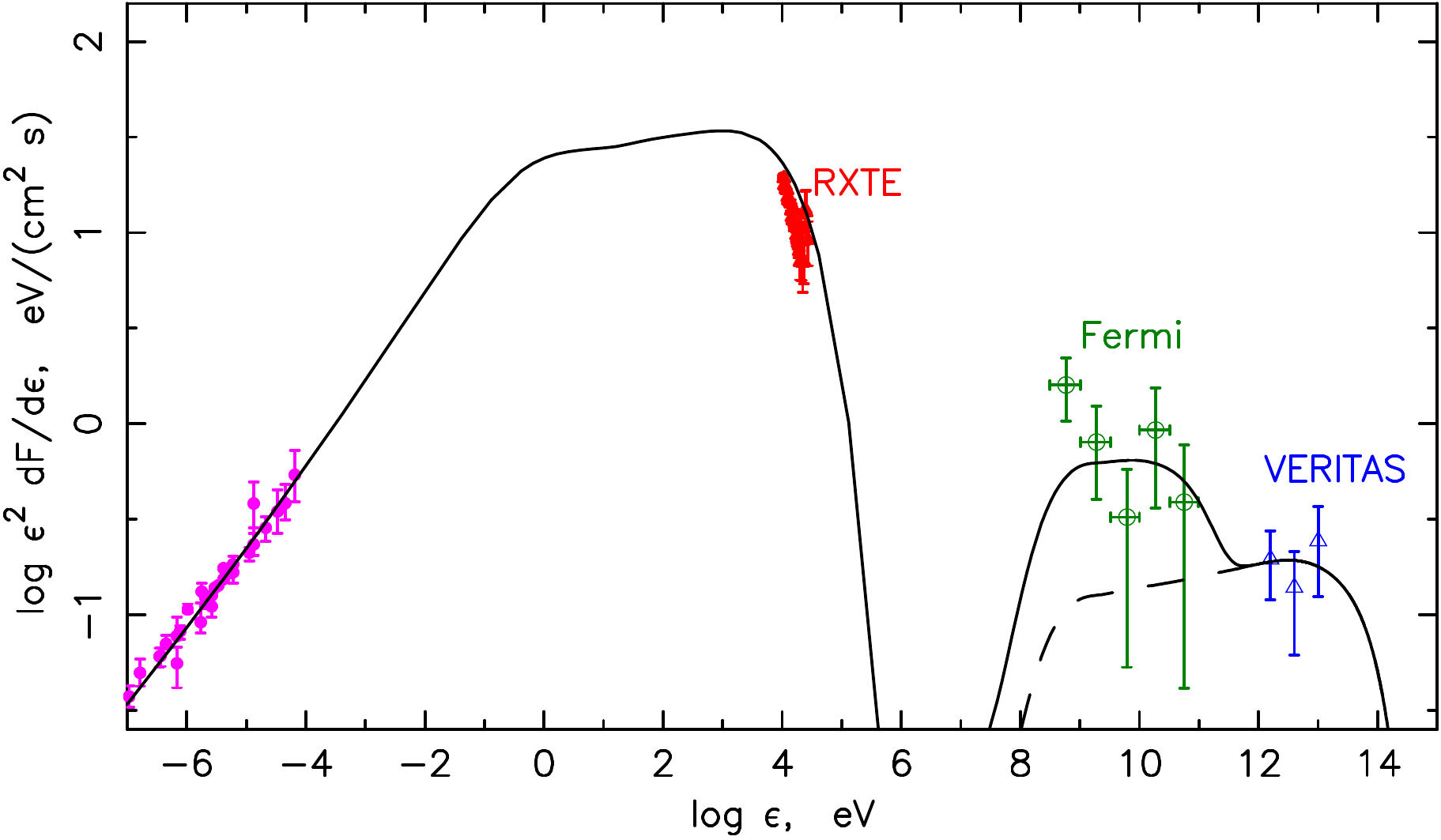}
\caption{Spatially integrated spectral energy distributions of  Tycho's SNR, 
calculated for a source
distance $d=3.8$~pc. The observed radio emission \citep{rey92},
X-ray \citep{agp99} and \gr emission data obtained
by  Fermi \citep{fermiTycho} and  VERITAS \citep{veritas11}
are shown as well.  Dashed and solid lines represent the
contribution of the warm-phase ISM and the total \gr energy spectrum that
includes the contribution of the clouds (see main text), respectively.}
\label{Fig6}
\end{figure}

The TeV \gr emission from Tycho detected by VERITAS \citep{veritas11}
  corresponds very well to the above expectation. As can be seen from
  Fig.\ref{Fig6} a new \gr spectrum calculated within the kinetic nonlinear
  theory (shown by the dashed line) is well consistent with the VERITAS
  measurements.  For the proton injection rate $\eta = 3\times10^{-4}$ this is
  still compatible with the above-mentioned shock modification and softening of
  the observed radio synchrotron emission spectrum. The new distance
  $d=3.8$~kpc and the corresponding new ambient ISM number density
  $N_\mathrm{g}=0.25$~cm$^{-3}$ were taken in order to fit the observed TeV \gr
  emission.

However the \gr spectrum measured by
the Fermi LAT at energies 400~MeV to 100~GeV \citep{fermiTycho} is considerably
(by a factor 2 to 5) above the value predicted by the kinetic theory (see
Fig.\ref{Fig6}).  This excess of GeV \gr emission, when compared with the
theoretical predictions, requires a more detailed consideration of this object
and its environment, taking into account new physical factors which had
been hitherto neglected.

The physics aspect which is not included in
the  spherically
symmetric model is an essential inhomogeneity of the ambient ISM on spatial
scales that are smaller than the SNR radius.  It is rather an inherent nonuniformity of
the average ISM on account of (i) the interplay between its radiative heating
by the diffuse galactic UV field and the radiative cooling of the gas and/or
(ii) the stochastic agitation of the ISM by the mechanical energy input from
supernova explosions occurring in the galactic environment. The first effect
is a thermal instability and thus a mechanism for small-scale cloud formation
in the ISM driven by runaway radiative cooling \citep{field1965}.
Specifically the balance between line-emission cooling and gas heating
due to the ultraviolet background radiation leads to two thermally stable
equilibrium ISM phases \citep{wolfire2003}. One of them is the so-called warm
interstellar medium with a typical gas number density  $N_\mathrm{g1} \sim
0.1$~cm$^{-3}$ and temperature
$T_1\approx 8000$~K, the other one a cold neutral medium with
$N_\mathrm{g2}\sim 10$~cm$^{-3}$ and $T_2\approx 100$~K.  According to
simulations the scale of dense clouds is typically $l_\mathrm{c}=0.1$~pc
(e.g. \citep{audit10}). Therefore the typical ISM should be
treated as two-phase medium composed of uniform warm ISM (phase I) and
small-scale dense clouds (phase II) embedded in the warm ISM. 

In order to determine the specific properties of the CRs and their
nonthermal emission in the case of a two-phase ISM the latter was approximated
in a simple form, as a uniform warm phase
with gas number density $N_\mathrm{g1}$ plus an ensemble of
small-scale dense clouds with gas number density $N_\mathrm{g2}$ \citep{bkv13}. The
warm diluted ISM phase is assumed to have a volume filling factor
$F_1\approx 1$, whereas the clouds occupy a relatively small fraction of
space with filling factor $F_2\ll 1$. It is in
addition assumed that most of the gas mass is contained in the warm phase,
which means that $F_1N_\mathrm{g1}\gg F_2N_\mathrm{g2}$.  Due to this fact the
SN shock propagates in the two-phase ISM without essential changes compared with
the case of a purely homogeneous ISM with number density
$N_\mathrm{g1}$. Therefore it produces
inside the phase~I of the ISM roughly the same amount of CRs and
nonthermal emission as in the case of a homogeneous ISM.  Then one has
to estimate the additional contribution of the clouds in order to
determine the overall spectrum of CRs.
Using the scaling laws for the spectrum amplitude 
and cutoff energy of shock accelerated CRs one can easily estimate the values
of relevant parameters of \gr emission created in shocked clumps
(amplitude and shape of \gr spectrum and its cutoff energy) and
determine the \gr flux $F_{\gamma2}(\epsilon_{\gamma})$ from the clumps.

The total \gr flux $F_{\gamma}=F_{\gamma1}+F_{\gamma2}$, expected from Tycho's
SNR for a two-phase ISM is shown in Fig.\ref{Fig6} \citep{bkv13}. One can see
that it fits the existing data in a satisfactory way. The considerable increase
(by a factor of 5) of the \gr emission at energies $\epsilon_{\gamma}< 100$~GeV
expected in the case of a purely homogeneous ISM is due to the contribution of
clumps which contain only 10\% of the ISM mass.

One can conclude that the observed properties of Tycho's SNR are consistent
with the efficient nuclear CR production even though this scenario requires the
confirmation that Tycho's SNR indeed expands into  a clumpy
ISM.

\subsection{Supernova remnant RX~J1713-3946}
RX~J1713-3946 is a  shell-type supernova remnant (SNR), located in
the Galactic plane, that was discovered in X-rays with ROSAT \citep{pfef96}.
The study of this SNR with the ASCA satellite \citep{koyama97,slane99} have
shown that the observable X-ray emission is entirely non-thermal, and this
property was confirmed in later XMM observations \citep{cassam04}.  The radio
emission of this SNR is weak: only part of the shell could be detected in radio
synchrotron emission up to now, with a poorly known spectral form
\citep{laz04}.

RX~J1713-3946 was also detected in very high energy \grs with the
CANGAROO \citep{muraishi00,enomoto02} and H.E.S.S.
\citep{aha06} 
telescopes. Especially the latter observations
show a clear shell structure at TeV energies which correlates well with the
ASCA contours.

The difficulty for the theoretical description is 
the fact that several key parameters of this 
source are either not known or poorly constrained. This already concerns 
the distance and age of the object. 
It was demonstrated that consistent description
of this object is achieved 
\citep{bv06} following present consensus which puts 
the distance at 1 kpc, the age to about 1600 years 
and that the primary explosion must have been a type 
II/Ib SN event with a massive progenitor star whose mass loss in the main 
sequence phase created a hot wind bubble in a high-density environment. 
The solution for the overall remnant dynamics then yields the value for 
the expansion velocity of the outer shock, given the total mechanical 
energy $E_{\mathrm{sn}}=1.8 \times 10^{51}$~erg 
released in the explosion. To obtain a consistent 
solution for the broadband nonthermal emission the injection rate $\eta=3\times 10^{-4}$
and
interior magnetic field $B_\mathrm{d}\approx 130~\mu$G are needed.
Note that since the value of radio spectral index $\alpha$ is not determined
the estimated proton injection rate is not as reliable as in the case of
SN~1006 or Tycho's SNR.

The properties of small scale structures of SNR \rxj seen in X-rays \citep{hiraga05}, provide even 
stronger evidence that the magnetic field inside the SNR is indeed considerably
amplified.

\begin{figure}
\includegraphics[width=0.47\textwidth]{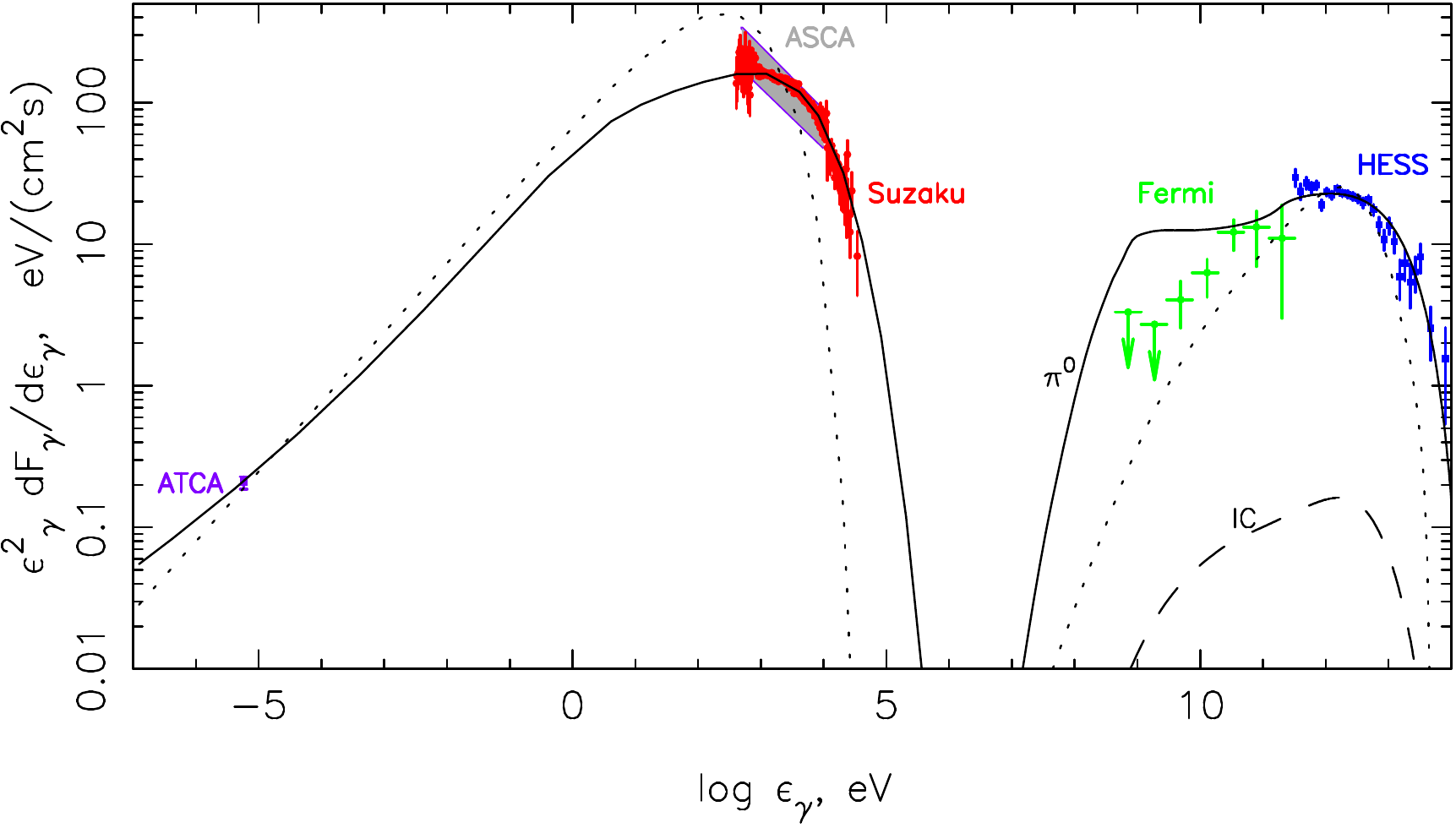}
\caption{Spatially integrated spectral energy distribution of \rxj \citep{bv10}. The 
{ATCA} radio data \citep[cf.][]{aha05}, {ASCA} X-ray data
\citep[cf.][]{aha06}, Fermi \citep{fermiRXJ1713} and
{H.E.S.S.} \citep{aha07} \gr data are
shown as well. The solid and dashed curves at
energies above $10^7$~eV correspond to $\pi^0$-decay and inverse Compton (IC) \gr emission, 
respectively.
The dotted line corresponds to the test
  particle limit which implies insignificant proton acceleration and magnetic
  field amplification.
}
\label{Fig7}
\end{figure}

The calculated overall broadband spectral energy distribution \citep{bv10} is displayed in
Fig.\ref{Fig7}, together with the experimental data from {ATCA} at radio
wavelengths, as estimated for the full remnant by \citet{aha06},
the X-ray data from {ASCA}, GeV \gr spectrum from Fermi \citep{fermiRXJ1713}
and the TeV \gr spectra
from  {H.E.S.S.} \citep{aha07}.
 The overall fit of the existing data except Fermi data, which appeared later on (see below),
is impressive, noting that the choice of a few
key parameters like $\eta$, $B_\mathrm{d}$, $E_{\mathrm{sn}}$ 
in the theory allows a spectrum determination over 
more than 19 decades.

The same observed spectrum can also be compared with a theoretical spectrum
(Fig.\ref{Fig7}) in which a very low ion injection rate ($\eta = 10^{-5}$) and a rather
low downstream magnetic field strength of $20~\mu$G was assumed. 
This corresponds to a dominantly leptonic \gr test particle spectrum
without field amplification. The
IC-scattered diffuse radiation field is the CMB without any additions. The
electron injection strength was fitted such that an optimum fit with the
observations in the radio and X-ray ranges is achieved, cf. Fig.\ref{Fig7}. We note that
the {\it form} of the X-ray spectrum is only very poorly fitted in this
leptonic scenario, since otherwise the radio data can not simultaneously be
fitted. Also the \gr spectrum has a maximum which is much too sharp in
comparison with the observed H.E.S.S. spectrum. At \gr energies of 1 GeV the
spectral energy flux density is a factor of about 30 below the value in the
hadronic scenario. Therefore it was expected that the measurements of
GeV-emission would provide the clear discrimination between these two scenarios.
However the recently appeared Fermi measurements of the \rxj spectrum at 
energies 1-100 GeV \citep{fermiRXJ1713} have not provided the clear resolution of this problem.
As it is seen in Fig.\ref{Fig7} Fermi data are situated almost in the middle
between the predictions of leptonic and hadronic scenarios.
While the hard measured spectrum at GeV energies is interpreted 
by the Fermi collaboration \citep{fermiRXJ1713}
and in some theoretical considerations \citep{za10, ellison12}
as a strong argument in favor of a leptonic scenario 
the origin of \gr emission of \rxj is still debated.

It was suggested, that the interpretation of the \gr observed emission of \rxj 
changes dramatically,
if the SNR expands in a clumpy medium \citep{za10,inoue12, ga14}. 
This is indeed expected if the SNR progenitor
is a massive star in a molecular cloud. The dense clumps remain
unshocked inside SNR if their density considerably (by a factor of about $10^5$)
exceeds the density ($N_\mathrm{g}\sim 10^{-2}$~cm$^{-3}$) of surrounding warm medium \citep{inoue12}.
CRs accelerated at the SN shock in the diluted medium penetrate the clumps
and efficiently produce there \gr emission.
Since CR diffusion coefficient is increasing function of energy,
higher energy CRs penetrate more efficiently. Therefore 
CR energy spectrum inside the clumps and corresponding 
spectrum of hadronic \gr emission are expected to be much harder compared with
the case of uniform medium. If the clumps contain the dominant fraction
of the mass in the SNR hadronic \gr emission can explain the
hard Fermi spectrum \citep{inoue12, ga14}.

The described scenario happens under the assumption that every
clump is surrounded by the transition layer containing magnetic field
which strength is highly enhanced (up to $\sim 100$~$\mu$G) compared
with its value in the surrounded medium. Such a layer
considerably reduce CR diffusion inside the clump so that
only CRs with energy above 1 TeV are able to penetrate the clump.
According to the results of MHD simulation \citep{inoue12}
this indeed can happen in the downstream region where
plasma flow speed shear produces magnetic field amplification
around the clump surface. 
However such an effect is expected to be much weaker upstream,
within the shock precursor region and downstream during
the initial time period $\Delta t \approx 0.1t$ (here $t$ is the SNR age)
when magnetic field is not yet amplified considerably \citep{inoue12}.
If the clumps in these
regions are not surrounded by the boundary layers,  CRs
more efficiently  penetrate the clumps 
compared to the simple estimates \citep{ga14}.
Then the \gr energy spectrum produced by these CRs 
is expected to be less harder 
as compared with estimates \citep{inoue12, ga14}.

We can conclude that further computational and observational efforts 
are needed in order to determine the nature
of \gr emission of \rxj.

\subsection{Supernova remnant  VelaJr.(RX J0852.0-4622) }
VelaJr.(or RX J0852.0-4622) is a shell-type SNR with a diameter of $2^{\circ}$,
located in the Galactic plane. The SNR was originally discovered in X-rays with
the ROSAT satellite \citep{aschen98}. In projection
along the line of sight, RX J0852.0-4622 lies entirely within the still much
larger Vela SNR and is only visible in hard X-rays, where the thermal radiation
from the Vela SNR is no longer dominant. While non-thermal emission from the
shell of RX J0852.0-4622 has been confirmed by several X-ray observatories 
\citep{slane01, bamba05}, a detection of thermal
X-ray emission from the shell or the interior is not well established because
of the confusion with the Vela SNR.

The radio emission of RX J0852.0-4622 is weak. In fact the SNR radio shell was only
identified  after
its discovery in X-rays. The radio spectrum of RX J0852.0-4622 is therefore not
well determined. Only for the northeastern rim a spectral index can be derived
with moderate accuracy \citep{duncang00}.

The shell of RX J0852.0-4622 was also detected in very high energy (VHE) \grs by the
H.E.S.S. collaboration \citep{aha05, aha07a}, with a \gr flux as large
as that from the Crab Nebula.  

Vela Jr. represents the SNR which is very similar to \rxj.
The difficulty for the theoretical description is 
the fact that  key parameters of this 
source (distance, age, radio emission spectrum)
are either not known or poorly constrained. 

\begin{figure}
\centering
\includegraphics[width=0.47\textwidth]{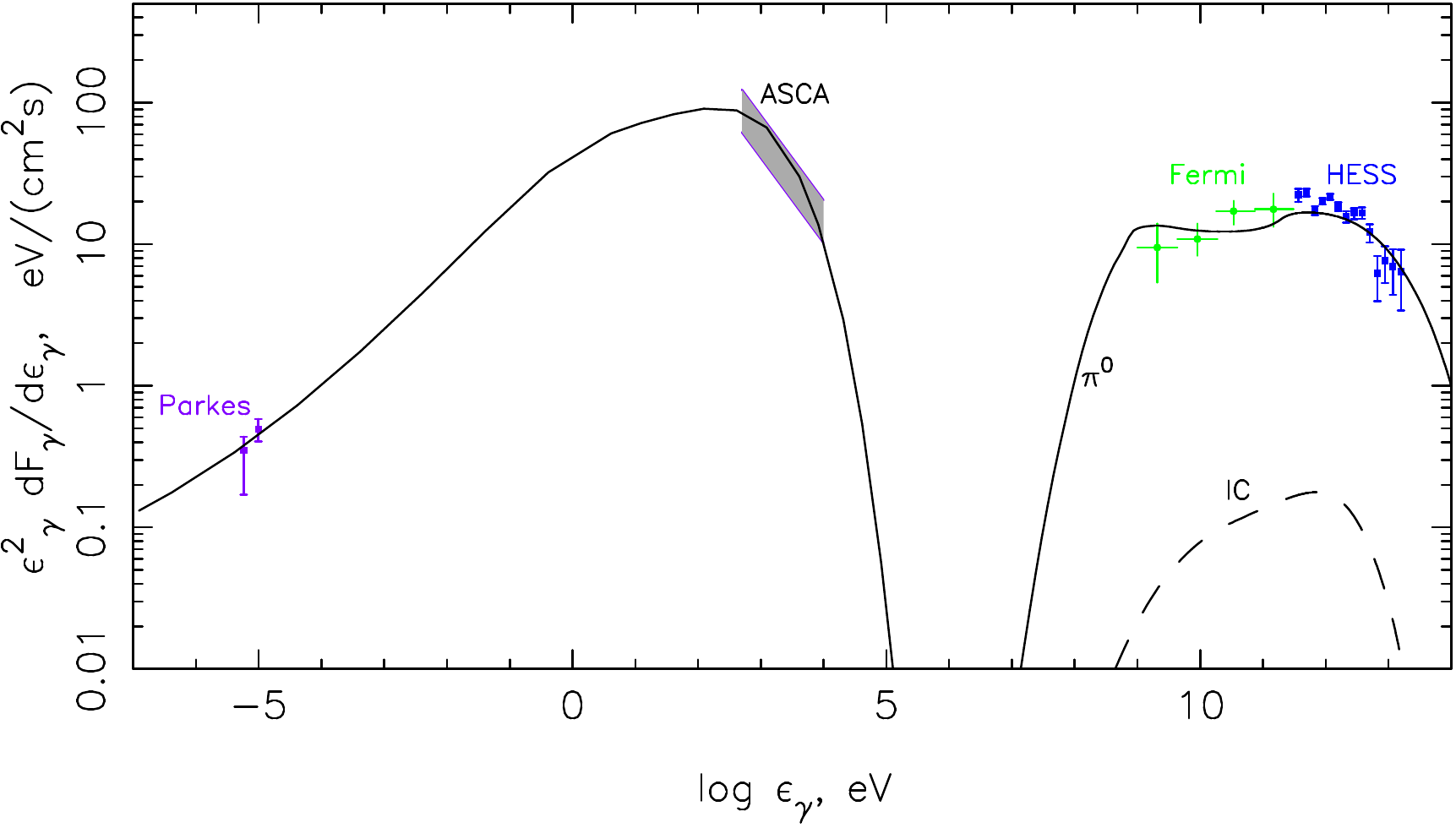}
\caption{Calculated broadband spectral energy densities of RX J0852.0-4622 
as functions of photon energy
$\epsilon_{\gamma}$.   Radio \citep{duncang00}
and X-ray synchrotron  fluxes from ASCA \citep{slane01}, ROSAT \citep{aschen98, aha07a} 
measurements  together with H.E.S.S. \cite{aha07a}
and Fermi \citep{fermiVelaJr} \gr data are shown as well.
}
\label{Fig8}
\end{figure}

The present-day parameters $d= 1$~kpc, $E_\mathrm{sn}=2\times 10^{51}$~ers, $t=3930$~yr,
$B_\mathrm{d}= 106$~$\mu$G, $\eta = 3\times 10^{-4}$ and $K_\mathrm{ep}\approx
2\times10^{-4}$, estimated from the overall fit of the observed
nonthermal emission,  lead to good agreement between the
calculated and the measured spectral energy distribution of the synchrotron
emission in the radio to X-ray ranges at the present time (Fig.\,\ref{Fig8}) \citep{bpv}. The
steepening of the electron spectrum at high energies due to synchrotron losses
and the smooth cutoff of the overall electron spectrum together naturally yield
a fit to the X-ray data with their soft spectrum. Such a smooth spectral
behavior is achieved in an assumed upstream field $20$~$\mu$G (which leads to
the above downstream field $B_\mathrm{d}$). 

Fig.\,\ref{Fig8} also shows the calculated \gr spectral energy distributions due
to $\pi^0$-decay, IC emission, and nonthermal Bremsstrahlung, together with the
existing experimental data.

According to the calculation, the hadronic $\gamma$-ray production exceeds the
electron contribution by more than two orders of magnitude at all energies. 
For energies
$\epsilon_{\gamma}=1-10$~TeV the \gr spectrum is expected to be close to
$dF_{\gamma}/d\epsilon_{\gamma}\propto \epsilon_{\gamma}^{-2}$, whereas
starting from $\epsilon_{\gamma}\approx 10$~TeV it has a smooth extended cutoff
despite the comparatively much sharper cutoff of the proton energy spectrum. 
It is clearly seen from Fig.\ref{Fig8} that
the calculated spectrum fits the H.E.S.S. measurements in a satisfactory way.

The predicted \gr emission \citep{bpv} at GeV-energies was very well confirmed
by the Fermi measurements \citep{fermiVelaJr}.  This leads to the conclusion
that RX J0852.0-4622 represents a hadronically dominated source of \grs and
typical source of GCRs.  However other considerations of this remnant argued
that the gas density required by hadronic models would imply an intense X-ray
line emission, which is not observed \citep{lee13}.  It is not clear at the
moment whether this problem could be eliminated by the 
adjustment of poorly constrained key parameters of RX J0852.0-4622 within
the hadronic model, which is so well consistent with observation.  Further
observational and theoretical studies are needed to obtain better estimates of
relevant astronomical SNR parameters and to obtain more justified conclusions
about the origin of \gr emission.

\subsection{Diffuse Galactic gamma-ray background radiation}
Since the source cosmic rays (SCRs) which are still confined  
inside all existing
active SNRs have much more hard spectrum compared with GCRs,
SCRs inevitably make a strong
contribution to the ``diffuse'' \gr flux from the Galactic disk at all
energies above a few GeV, if the population of SNRs is the main
source of the GCRs. According to estimates \citep{bv00b}, 
the SCR contribution
becomes significant at energies greater than 100~GeV.

The diffuse Galactic \gr emission (DGE) from the full sky has recently been
analyzed and compared with the observations with the 
{\it Fermi}-LAT for high energies  $200~\mathrm{MeV} \leq
\epsilon_{\gamma} \leq 100~\mathrm{GeV}$ \citep{ackermann12}. The DGE had been
modeled using the GALPROP code (e.g. \citep{strong07}). 
These phenomenological models were constrained to reproduce
  directly measured GCR data and were then used to
  calculate the DGE. To construct a model for the
  expected total \gr emission, the \gr emission from the resolved point sources
  together with the residual instrumental \gr background and the extragalactic
  diffuse \gr background -- both assumed to be isotropic --
  were added to the DGE model. In the inner Galaxy, the emission of the
  resolved sources apparently reaches a fraction of $\sim 50$~percent of the
  expected overall spectral energy flux density at $\epsilon_{\gamma}\approx
  100$~GeV \citep{ackermann12}.

  In the Galactic Plane these
  models systematically underpredict the data above a few GeV, and they do so
  increasingly above about 10 GeV until 100 GeV \citep{ackermann12}. 
In the present paper this difference between data and
  model will be called the "{\it Fermi}-LAT Galactic Plane Surplus"
  (FL-GPS). It is most pronounced in the inner Galaxy. According to
  \citet{ackermann12}, it can however also be seen in the outer Galaxy, with
  even a small excess at intermediate latitudes.

\begin{figure}
  \includegraphics[width=0.47\textwidth]{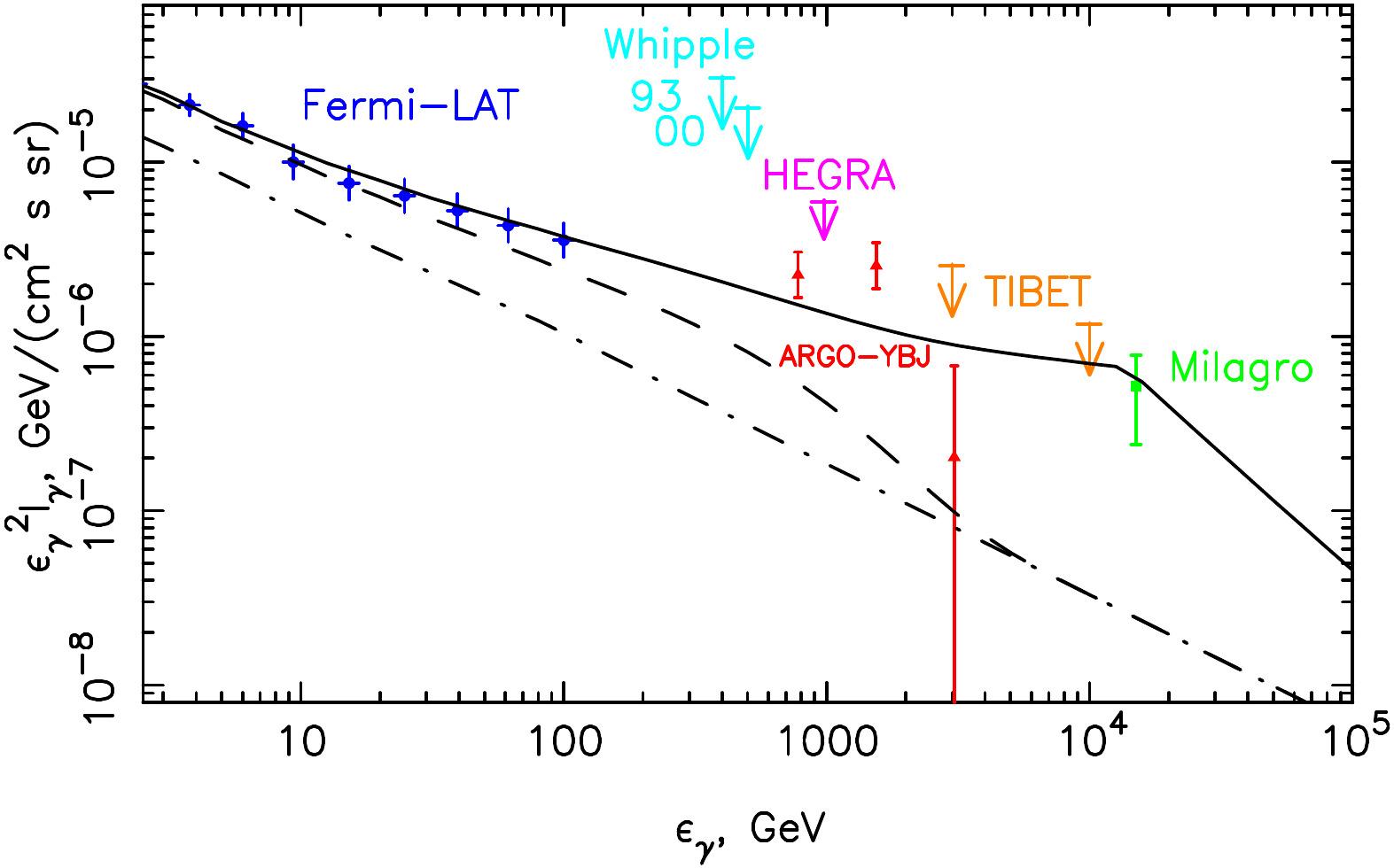} 
\caption{ The diffuse \gr spectrum of the low
    latitude inner Galaxy ($-80^{\circ}<l<80^{\circ}$, $|b|\leq8^{\circ}$).
    The expected $\pi^0$-decay \gr component $I_\mathrm{GCR}^{\pi}$ created by
    truly diffuse GCRs in the Galactic Disk is shown by the dash-dotted
    line. The total expected emission $I_\mathrm{tot}$ (= modeled diffuse
    Galactic emission $I_\mathrm{DGE}$ plus detected sources and isotropic
    backgrounds) up to $\epsilon_{\gamma} = 100$~GeV \citep{ackermann12}
    corresponds to the dashed line. This $I_\mathrm{tot}$ is extrapolated
    beyond 100 GeV by a power law with an assumed cutoff in
    $I_\mathrm{tot}-I_\mathrm{GCR}^{\pi}$. The solid line represent
    the total expected emission including the contribution of SCRs confined
    inside unresolved SNRs. The {\it Fermi}-LAT \citep{ackermann12},
 ARGO-YBJ  \citep{ma11},
    Milagro  \citep{milagro} data and  Wipple \citep{reyn93,leboh00},  HEGRA \citep{aha01}
    and Tibet \citep{tibet} upper limits   are shown  as well.
\label{Fig9} }
\end{figure}

The modern gamma-ray instruments are able to detect
individual SNR situated at relatively small distance
$d\lsim 1$~kpc. Therefore most of the existing Galactic SNRs
can not be resolved and contribute in DGE only.
In Fig.\ref{Fig9} a calculated \gr spectrum
$I_{\gamma}=I_\mathrm{tot}+I_\mathrm{SCR}$ of the low latitude inner Galaxy
($-80^{\circ}<l<80^{\circ}$, $|b|\leq8^{\circ}$) is presented \citep{bv13}. Besides the
overall emission model $I_\mathrm{tot}$ 
it includes the contribution $I_\mathrm{SCR}$ of SCRs
confined within unresolved SNRs. 

As shown in Fig.\ref{Fig9}, the discrepancy between the observed
``diffuse'' intensity and standard model predictions at energies above a few
GeV \citep{ackermann12} can be attributed to the SCR contribution alone up to
$\sim 100$~GeV.

As one can see further from Fig.\ref{Fig9}, the expected \gr flux at
$\epsilon_{\gamma}=1$~TeV is consistent with the HEGRA upper limit, the Whipple
upper limits \citep{reyn93}, the Tibet-AS upper limits \citep{tibet}, and with
the fluxes measured by the Milagro detector \citep{milagro} at
$\epsilon_{\gamma}=15$~TeV, as well as with those obtained
with the ARGO-YBJ detector \citep{ma11} in the TeV range, taken at face
value. Unfortunately, existing measurements do not coincide well
  in their longitude (l) / latitude (b) coverage with the corresponding ranges
  for {it Fermi}-LAT that correspond to $-80^{\circ}<l<+80^{\circ}$ /
  $-8^{\circ}<b<+8^{\circ}$. For this reason the following comparisons must be
  taken with reservations.

These considerations demonstrate that the SCRs inevitably make a strong
contribution to the ``diffuse'' \gr flux from the Galactic disk at all energies
above a few GeV, if the population of shell-type SNRs is the main source of the
GCRs. This explains the {\it Fermi}-LAT Galactic Plane Surplus. The
quantitative estimates show  in addition that the SCR contribution
dominates over the extrapolated Fermi-LAT model for the total diffuse emission
- which includes detected sources and isotropic Backgrounds -  at
energies greater than 100~GeV due to its substantially harder spectrum. The
diffuse emission measured by Milagro at 15~TeV and by ARGO-YBJ at TeV
  energies provide limited evidence for that.  

\section{Conclusions}
The proton GCR energy spectrum measured in ATIC-2, AMS 02, PAMELA and KASKADE
experiments is generally consistent with its origin in Galactic SNRs.
It has a slightly concave shape and extends up to a maximal energy
$\epsilon_\mathrm{max}\approx 2\times 10^{15}$~eV as expected from nonlinear
theory of CR acceleration in SNRs.

ATIC-2 data alone clearly show that at energies 30~GeV -- 30~TeV
Helium energy spectrum is essentially harder then the proton spectrum.
Considering ATIC-2 data together with PAMELA and AMS 02 data it is
already not so obvious that Helium energy spectrum is indeed so much
harder at energies above 100~GeV.

SNR contribution into the secondary CR spectra 
and into the diffuse \gr background represents the 
hard component
which is unavoidably expected if SNRs are the main source of GCRs.
Comparison of the data, obtained in RUNJOB, Fermi, PAMELA and AMS 02 experiments, 
with theoretical expectations leads to a conclusion 
that the observed high energy excess of positrons, secondary nuclei
and diffuse \gr background
can be  
produced in Galactic SNRs. This enable to expect  similar
excess in the antiproton energy spectrum. The data expected very soon from AMS-02
experiment will make it clear whether the actual ratio  $\bar p/p$
is indeed not less flat at energies $\epsilon_\mathrm{k}> 10$~GeV then it is expected
due to SNRs contribution.

Existing SNR emission data 
provide the evidences for the
efficient CR production in SNRs accompanied by significant magnetic field
amplification. At the same time the situation here is not completely clear yet:
together with the cases of SN~1006 and Tycho's SNR which observed properties
well correspond to the theoretical expectations, there are other young SNRs
\rxj and  VelaJr.(RX J0852.0-4622) for  which the nature of the detected
\gr emission 
is difficult to determine 
mainly because some  key SNR parameters are not known or poorly constrained.

The further progress in our understanding of CR origin will be achieved
due to the new measurements of GCR spectra and SNR emission spectra
performed with essentially more sensitive instruments.

\section*{Acknowledgments}
The author thanks H.J. V\"olk for valuable comments.
This work is supported by the Russian Foundation for Basic Research (grants
13-02-00943 and 13-02-12036) and by the Council of the
President of the Russian Federation for Support of Young Scientists and Leading
Scientific Schools (project No. NSh-3269.2014.2).


\nocite{*}
\bibliographystyle{elsarticle-num}
\bibliography{martin}

\begin{thebibliography}{00}


\bibitem[Ginzburg and Syrovatski(1964)]{gs64}
V.L. Ginzburg, S.I. Syrovatskii, {The Origin of Cosmic Rays},
Pergamon Press and Macmillan Comp. 1964.
\bibitem[Berezinsky et al.(1990)]{berezinetal90}
V.S. Berezinsky, S.V. Bulanov, V.L. Ginzburg et al.,
{Astrophysics of Cosmic Rays}, North-Holland Publ. Comp.1990.

\bibitem{krym77}
G.F. Krymsky, Soviet Phys. Dokl. 23 (1977) 327.

\bibitem[Drury(1983)]{drury83}
L'O.C. Drury,  
Rep. Progr. Phys. 46 (1983) 973.

\bibitem[Blandford and Eichler(1987)]{blei87}
R. Blandford, D. Eichler, Phys. Rep. 154 (1987) 1.

\bibitem[Berezhko and Krymsky(1988)]{bk88}
E.G. Berezhko, G.F. Krymsky,  
Soviet Phys.-Uspekhi 31 (1988) 27.

\bibitem[Malkov and Drury(2001)]{dm}
M.A. Makov, L'O.C. Drury, Rep. Progr. Phys. 64 (2001) 429.

\bibitem[Berezhko(2008)]{ber08}
 E.G. Berezhko, Adv. Space Res. 41 (2008) 429.

\bibitem[Berezhko et al.(1996)]{bek96} 
E.G. Berezhko et al., J. Exp. Theor.
Phys. 82 (1996) 1.

\bibitem[Berezhko and V\"olk(1997)]{bv97}
E.G. Berezhko, H.J. V\"olk,  Astropart. Phys. 7 (1997) 183.

\bibitem[Berezhko and V\"olk(2000a)]{bv00a}
E.G. Berezhko, H.J. V\"olk, Astron. Astrophys. 357 (2000) 283.

\bibitem[Zirakashvili and Ptuskin(2009)]{zp09}
V.N. Zirakashvili, V.S. Ptuskin, in: Proc. 31st ICRC, Lodz 2009.

\bibitem[Kang(2010)]{kang10}
H. Kang, J. Korean Astron. Soc. 43 (2010) 25.

\bibitem[Berezhko et al.(2003)]{bkpvz03}
E.G. Berezhko et al., Astron. Astrophys. 410 (2003) 189.

\bibitem[Lucek \& Bell(2000)]{lb00}
S.G. Lucek, A.R. Bell, MNRAS 314 (2000) 65.

\bibitem[Bell \& Lucek(2001)]{belll01}
A.R. Bell, S.G. Lucek, MNRAS 321 (2001) 433.

\bibitem[Bell(2004)]{bell04}
A.R. Bell, MNRAS 353 (2004) 550. 

\bibitem[Caprioli(2014)]{caprioli14}
D. Caprioli, A. Spitkovsky, arXiv:1401.7679v1[astro-ph.HE]

\bibitem[{{Bamba} {et~al.}(2003){Bamba}, {Yamazaki}, {Ueno}, \&
  {Koyama}}]{bamba03}
A. {Bamba} et al., Astrophys. J. 589 (2003) 827; 2014, these proceedings.

\bibitem[Long et al.(2003)]{long03}
K.S. Long et al. , Astrophys. J. 586 (2003) 1162.

\bibitem[Vink \& Laming(2003)]{vl03}
J. Vink, J.M. Laming, Astrophys. J. 548 (2003) 758

\bibitem[{{Berezhko} {et~al.}(2003){Berezhko}, {Ksenofontov}, \& {V{\" o}lk}}]{bkv03} 
E.~G. {Berezhko} et al., Astron. Astrophys. 412 (2003) L11

\bibitem[{{Berezhko} \& {V{\" o}lk}(2004)}]{bv04}
E.~G. {Berezhko}, H.~J. {V{\" o}lk}, Astron. Astrophys. 419 (2004) L27.

\bibitem[Reville \& Bell(2012)]{reville12}
B. Reville, A.R. Bell, MNRAS 419 (2012) 2433.

\bibitem[{V{\" o}lk} et al.(2005)]{vbk05}
H.~J. {V{\" o}lk} et al., A\&A 505 (2005) 169.

\bibitem[Berezhko(1996)]{ber96}
E.G. Berezhko, Astropart. Phys. 5 (1996) 367.

\bibitem[Berezhko \& V\"olk(2007)]{bv07}
E.G. Berezhko, H. J. V\"olk,  Astrophys. J. 661 (2007) L175.

\bibitem[Ptuskin et al.(2010)]{pzs10}
V.S. Ptuskin, Astrophys. J. 718 (2010) 31.



\bibitem[Boezio et al.(2003)]{caprice}
M. Boezio et al.,  Astropart. Phys. 19 (2003) 583.

\bibitem[AMS (2013)]{ams2}
http://www.ams02.org

\bibitem[Pamela (2009)]{pamelaCR}
O. Adriani et al., Science 332 (2011) 69.

\bibitem[Panov et al.(2006)]{atic2}
A. D. Panov et al. 2006, astro-ph/0612377.

\bibitem[Asakimori et al.(2003)]{jacee}
K. Asakimori et al.  Astrophys. J. 502 (2003) 278.

\bibitem[Antoni et al.(2005)]{kascade}
T. Antoni et al.  Astropart. Phys. 24 (2005) 1.

\bibitem[Berezhko and Ksenofontov (2014)]{bk14}
E.G. Berezhko, L.T. Ksenofontov, Astrophys. J. 791 (2014) L22.

\bibitem[Berezhko et al.(2012)]{bkv12}
E.G. Berezhko et al., Astrophys. J. 759 (2012) 12.

\bibitem[Berezhko et al.(2009)]{bkv09}
E.G. Berezhko et al., A\&A 505 (2009) 169


\bibitem[Adriani et al.(2010)]{pamela10}
O. Adriani et al.,  Phys. Rev. Lett. 105 (2010) 121101.

\bibitem[Donato et al.(2001)]{donato01}
F. Donato et al.,  Astrophys. J. 563 (2001) 172.

\bibitem[Blasi \& Serpico(2009)]{BlasiSer09}
P. Blasi, P.D. Serpico,  Phys. Rev. Lett. 103 (2009) 081103.

\bibitem[Engelmann et al.(1990)]{heao3}
J.J. Engelmann et al., Astron. Astrophys. 233 (1990) 96.

\bibitem[Panov et al.(2008)]{atic2bc}
A.D. Panov et al., in: Proc. of 30th ICRC, Merida 2007, 
V. 2, p. 3.

\bibitem[Ahn et al.(2008)]{cream}
H.S. Ahn et al.,  Astropart. Phys. 30 (2008) 133.

\bibitem[Derbina et al.(2005)]{runjob}
V.A. Derbina et al., Astrophys. J. 628 (2005) L41.

\bibitem[Aguilar(2013)]{ams02}
M. Aguilar,  CERN Courier 53 (2013) 22.
 
\bibitem[Aguilar et al.(2007)]{ams01}
M. Aguilar et al., Phys. Lett. B 646 (2007) 145.

\bibitem[Beatty et al.(2004)]{heat}
J.J. Beatty et al., Phys. Rev. Lett. 93 (2004) 241102.

\bibitem[Adriani et al.(2009)]{pamela09}
O. Adriani et al. Nature 458 (2009) 607.

\bibitem[Ackermann et al.(2012)]{fermi12}
M. Ackermann et al., Phys. Rev. Lett. 108 (2012) 011103.

\bibitem[Aguilar et al.(2013)]{ams13}
M. Aguilar et al., Phys. Rev. Lett. 110 (2013) 141102.

\bibitem[Berezhko \& Ksenofontov(2013)]{bk13} 
E.G. Berezhko, L.T. Ksenofontov, J. Phys.: Conf. Ser. 409 (2013) 012025.

\bibitem[Moskalenko \& Strong(1998)]{ms98}
I.V. Moskalenko, A.W. Strong, Astrophys. J., 493 (1998) 694.

\bibitem[Drury et al.(1994)]{druryetal94}
L.O'C. Drury et al., Astron. Astrophys. 287 (1994) 959.

\bibitem[Allen et~al.(2004)]{allen04}
G.E. Allen et al., Adv. Space Res. 33 (2004) 440. 

\bibitem[Bamba et~al.(2008)]{bamba08}
A. Bamba et al., PASJ 60 (2008) S153.

\bibitem[Acero et al.(2010)]{sn1006hess10}
F. Acero et al., Astron. Astrophs. 516 (2010) A62.

\bibitem[Araya (2012)]{sn1006GeV}
M. Araya, F. Frutos, MNRAS 425 (2012) 2810.

\bibitem[Ptuskin \& Zirakashvili(2003)]{pz03}
V.~S. Ptuskin, V.~N. Zirakashvili, A\&A 403 (2003) 1.

\bibitem[V\"olk et al.(2008)]{vbk08}
 H.J. V\"olk et al., A\&A 483 (2008) 529.

\bibitem[Badenes et al.(2006)]{bad06}
C. Badenes et al., Astrophys. J. 645 (2006) 1373.

\bibitem[Reynolds \& Ellison(1992)]{rey92}
S.P. Reynolds, D.C. Ellison, Astrophys. J. 399 (1992) L75.

\bibitem[Acciari et al.(2011)]{veritas11}
V.~A. Acciari et al., Astrophys. J. 730 (2011) L20.

\bibitem[Allen et al.(1999)]{agp99}
G.E. Allen et al., in:  
26th ICRC, Salt Lake City 1999, V.3, p.480.

\bibitem[Giordano et al.(2012)]{fermiTycho}
F. Giordano, et al., Astrophys. J. 744 (2012) L2.

\bibitem[Field(1965)]{field1965}
G.~B. Field, Astrophys. J. 142 (1965) 531.

\bibitem[Wolfire et al.(2003)]{wolfire2003}
M.~G. Wolfire et al., 
Astrophys. J. 587 (2003) 278.

\bibitem[Audit \& Hennebelle(2010)]{audit10}
E. Audit, P. Hennebelle, A\&A 511 (2010) A76.

\bibitem[Berezhko et al.(2013)]{bkv13}
E.G. Berezhko et al., Astrophys. J. 763 (2013) 14.

\bibitem[Pfeffermann \& Aschenbach(1996)]{pfef96}
E. Pfeffermann, B. Aschenbach, 1996, in Roentgenstrahlung from the Universe,
ed. H.H. Zimmermann, J.  Tr\"umper, \& H. Yorke (MPE Rep. 263; Garching: MPE),
267.

\bibitem[Koyama et al.(1997)]{koyama97}
K. Koyama et al. PASJ 49 (1997) L7.

\bibitem[Slane et al.(1999)]{slane99}
P. Slane et al., Astrophys. J. 525 (1999) 357.

\bibitem[Cassam-Chena\"i et al.(2004)]{cassam04}
G. Cassam-Chena\"i et al., A\&A 427 (2004) 199.

\bibitem[Lazendic et al.(2004)]{laz04}
J.S. Lazendic et al., Astrophys. J. 602 (2004) 271.

\bibitem[[Muraishi et al.(2000)]{muraishi00}
H. Muraishi et al., A\&A 374 (2000) 895.

\bibitem[Enomoto et al.(2002)]{enomoto02}
R. Enomoto et al., Nature 416 (2002) 823.

\bibitem[Aharonian et al.(2004)]{aha04}
F.A. Aharonian et al., Nature 432 (2004) 75.

\bibitem[Aharonian et al.(2007)]{aha07}
F.A. Aharonian et al.,  A\&A 464 (2007) 235.

\bibitem[Aharonian et al.(2006)]{aha06}
F.A. Aharonian et al.,  A\&A 449 (2006) 223.

\bibitem[Berezhko and V\"olk(2006)]{bv06}
E.G. Berezhko, H.J. V\"olk, A\&A 451 (2006) 981.

\bibitem[Hiraga et al.(2005)]{hiraga05}
J.S. Hiraga et al., A\&A 431
(2005) 953.

\bibitem[Berezhko and V\"olk(2009)]{bv10}
E.G. Berezhko, H.J. V\"olk, A\&A 511 (2010) A34.

\bibitem[fermi (2011)]{fermiRXJ1713}
A.A. Abdo et al., Astrophys. J. 734 (2011) 28.

\bibitem[Zirakashvili, Aharonian(2010)]{za10}
V.N. Zirakashvili, F.A. Aharonian, Astrophys. J.708 (2010) 965.

\bibitem[Ellison et al.(2012)]{ellison12}
D.C. Ellison et al., Astrophys. J. 744 (2012) 39.

\bibitem[Inoue et al.(2012)]{inoue12}
T. Inoue et al. Astrophys. J. 744 (2012) 71.

\bibitem[Gabici and Aharonian(2014)]{ga14}
S. Gabici, F.A. Aharonian, arXiv:1406.2322v1 [astro-ph.HE].

\bibitem[Aschenbach(1998)]{aschen98}
B. Aschenbach, Nature 396 (1998) 141.

\bibitem[Slane et al.(2001)]{slane01}
P. Slane et al., Astrophys. J. 548 (2001) 814.

\bibitem[Bamba et al.(2005)]{bamba05}
A. Bamba et al., Astrophys. J. 632 (2005) 294.

\bibitem[Duncan \& Green(2000)]{duncang00}
A.R. Duncan, D.A. Green, A\&A 364 (2000) 732.

\bibitem[Aharonian et al.(2005)]{aha05}
 F.A. Aharonian et al.,A\&A 437 (2005) L7.

\bibitem[Aharonian et al.(2007)]{aha07a}
F.A. Aharonian et al., Astrophys. J. 661 (2007) 236. 

\bibitem[Fermi (2011)]{fermiVelaJr}
T. Tanaka et al.,, Astrophys. J. 740 (2011) L41.

\bibitem[Berezhko et al.(2009)]{bpv}
E.G. Berezhko et al., Astron. Astrophys. 505 (2009) 641.

\bibitem[Lee et al.(2012)]{lee13}
S.H. Lee et al., Astrophys. J. 767 (2013) 20.

\bibitem[Berezhko and V\"olk(2000b)]{bv00b} 
E.G. Berezhko, H.J. V\"olk,  Astrophys. J. 540 (2000) 
923. 

\bibitem[Ackermann et al.(2012)]{ackermann12}
M. Ackermann et al., Astrophys. J. 750 (2012) 3.

\bibitem[Strong (2007)]{strong07}
A.W. Strong, AApSS 309 (2007) 35.

%
%

\bibitem[V\"olk and Berezhko(2013)]{bv13}
H.J. V\"olk, E.G. Berezhko, Astrophys. J. 777 (2013) 149.

\bibitem[Reynolds et al.(1993)]{reyn93}
P.T. Reynolds et al., Astrophys. J. 404 (1993) 206.

\bibitem[Amenomori et al.(2005)]{tibet}
M. Amenomori et al.,  in: Proc. 29th ICRC 
(Pune, 2005) 4, 43

\bibitem[Abdo et al.(2009a)]{milagro} 
A.A. Abdo et al., Astrophys. J. 688 (2008) 1078.

\bibitem[Ma et al.(2011)]{ma11}
L.L. Ma et al., in: Proc. 32nd ICRC, Beijing 2011, doi: 10.7529/ICRC2011/V07/0256.

\bibitem[Aharonian et al.(2001)]{aha01} 
F.A. Aharonian et al., A\&A 
375 (2001) 1008. 

\bibitem[LeBohec et al.(2000)]{leboh00}
S. LeBohec et al., Astrophys. J. 539 (2000) 209.


\end{thebibliography}



\end{document}